\pdfoutput=1
\documentclass[a4paper,fleqn,usenatbib]{mnras}

\usepackage{txfonts}
\usepackage[T1]{fontenc}
\usepackage{ae,aecompl}

\usepackage{graphicx}	% Including figure files
\title[Long-Term Photometry of IC 348]{Long-Term Photometry of IC 348 with the YETI Network}

\author[D. J. Fritzewski et al.]{D.~J.~Fritzewski,$^{1,2}$\thanks{E-mail: dfritzewski@aip.de}
M.~Kitze,$^{1}$
M.~Mugrauer,$^{1}$
R.~Neuhäuser,$^{1}$
\newauthor
C.~Adam,$^{1}$
C.~Briceño,$^{3}$
S.~Buder,$^{1,4}$
T.~Butterley,$^{5}$
W.-P.~Chen,$^{6}$
B.~Din\c{c}el,$^{1,7}$
\newauthor
V.~S.~Dhillon,$^{8,9}$
R.~Errmann,$^{1,10}$
Z.~Garai,$^{11}$
H.~F.~W.~Gilbert,$^{1}$
C.~Ginski,$^{1,12}$
\newauthor
J.~Greif,$^{1}$
L.~K.~Hardy,$^{8}$
J.~Hern\'{a}ndez,$^{13,14}$
P.~C.~Huang,$^{6}$
A.~Kellerer,$^{15}$
E.~Kundra,$^{11}$
\newauthor
S.~P.~Littlefair,$^{8}$
M.~Mallonn,$^{2}$
C.~Marka,$^{1,16}$
A.~Pannicke,$^{1}$
T.~Pribulla,$^{11}$
St.~Raetz,$^{1,17}$
\newauthor
J.~G.~Schmidt,$^{1,18}$
T.~O.~B.~Schmidt,$^{1,19}$
M.~Seeliger,$^{1}$
R.~W.~Wilson,$^{5}$
and V.~Wolf$^{1}$
\\
$^{1}$Astrophysikalisches Institut und Universitäts-Sternwarte Friedrich-Schiller-Universität Jena, Schillergässchen 2-3, 07745 Jena, Germany\\
$^{2}$Leibniz-Institut für Astrophysik (AIP), An der Sternwarte 16, 14482 Potsdam, Germany\\
$^{3}$Cerro Tololo Interamerican Observatory, Casilla Postal 603, La Serena 1700000, Chile\\
$^{4}$Max-Planck-Institut für Astronomie, Königstuhl 17, 69117 Heidelberg, Germany\\
$^{5}$Centre for Advanced Instrumentation, Department of Physics, University of Durham, South Road, Durham DH1 3LE, UK\\
$^{6}$Graduate Institute of Astronomy, National Central University, 300 Jhongda Road, Jhongli 32001, Taiwan\\
$^{7}$Institut für Astronomie und Astrophysik Tübingen, Sand 1, 72076 Tübingen, Germany\\
$^{8}$Department of Physics and Astronomy, University of Sheffield, Sheffield S3 7RH, UK\\
$^{9}$Instituto de Astrofìsica de Canarias, 38205 La Laguna, Tenerife, Spain\\
$^{10}$Abbe School of Photonics, Max-Wien-Platz 1, 07743 Jena, Germany\\
$^{11}$Astronomical Institute, Slovak Academy of Sciences, 059 60 Tatransk\'{a} Lomnica, Slovakia\\
$^{12}$Leiden Observatory, Leiden University, PO Box 9513, 2300 RA Leiden, the Netherlands\\
$^{13}$Centro de Investigaciones de Astronom\'{i}a, Apdo. Postal 264, M\'{e}rida 5101-A, Venezuela\\
$^{14}$Instituto de Astronom\'{i}a, Universidad Nacional Aut\'{o}noma de M\'{e}xico, Unidad Acad\'{e}mica en Ensenada, Ensenada 22860, M\'{e}xico\\
$^{15}$Battcock Centre for Experimental Astrophysics, Cavendish Laboratory, Cambridge University, JJ Thompson Avenue,\\ Cambridge, CB3 0HE, UK\\
$^{16}$Instituto Radioastronomía Milimétrica (IRAM), Avenida Divina Pastora 7, E-18012 Granada, Spain\\
$^{17}$ESTEC - SCI-S ESA - European Space Agency, Keplerlaan 1, 2201 AZ Noordwijk, The Netherlands\\
$^{18}$Theoretische Astrophysik, Universität Tübingen, Auf der Morgenstelle 10,
72076 Tübingen, Germany\\
$^{19}$Hamburger Sternwarte, Gojenbergsweg 112, 21029 Hamburg, Germany\\
}

\date{Accepted XXX. Received YYY; in original form ZZZ}

\pubyear{2016}

\begin{document}
\label{firstpage}
\pagerange{\pageref{firstpage}--\pageref{lastpage}}
\maketitle

\begin{abstract} %max 250 words
We present long-term photometric observations of the young open cluster IC~348 with a baseline time-scale of 2.4\,yr. Our study was conducted with several telescopes from the Young Exoplanet Transit Initiative (YETI) network in the Bessel $R$ band to find periodic variability of young stars. We identified 87 stars in IC~348 to be periodically variable; 33 of them were unreported before. Additionally, we detected 61 periodic non-members of which 41 are new discoveries. Our wide field of view was the key to those numerous newly found variable stars. The distribution of rotation periods in IC~348 has always been of special interest. We investigate it further with our newly detected periods but we cannot find a statistically significant bimodality. We also report the detection of a close eclipsing binary in IC~348 composed of a low-mass stellar component ($M \gtrsim 0.09\,\mathrm{M}_{\sun}$) and a K0 pre-main sequence star ($M \approx 2.7\,\mathrm{M}_{\sun}$). Furthermore, we discovered three detached binaries among the background stars in our field of view and confirmed the period of a fourth one.
\end{abstract}


%%%%%%%%%%%%%%%%%%%%%%%%%%%%%%%%%%%%%%%%%%%%%%%%%%

%%%%%%%%%%%%%%%%% BODY OF PAPER %%%%%%%%%%%%%%%%%%

\section[]{Introduction}
\label{sec:intro}

The transit technique for finding planetary candidates is applied in several ground-based (e.g. \citealt{hatp} and  \citealt{mearth}) and space-based surveys (e.g. \citealt{corot} and \citealt{TESS}). This work is part of a project within the  ground-based Young Exoplanet Transit Initiative (YETI) \citep{NeuhauserYETI}. We used the data obtained by this survey to search for general periodic variation in the young open cluster IC~348. For an overview on the current state of YETI we refer the reader to the recent work of \cite{garaiYETI}.

In this paper we focus on the young (2\,Myr, \citealt{luhmanMembers}), nearby (316\,pc, \citealt{herbig}), and compact ($D\sim 20\arcmin$, \citealt{luhmanMembers}) open cluster IC~348. It is an intensely studied region of ongoing star formation. The T-Tauri stars (TTS) found therein by \cite{HerbigTTS} played an important role in the exploration of star formation \citep{herbig}. The  theory that TTS are young was strengthened and later confirmed through colour measurements and theoretical evolutionary models with the help of those TTS.

It is well known that IC~348 is younger than 10\,Myr but its age and the age-spread are still under debate. \cite{luhmanIMF} found an age spread from 0.5\,Myr to 10\,Myr from the photometric scatter in the colour-magnitude diagram. A similar wide age-spread was assumed in the Orion Nebula Cluster but it was shown by \cite{jeffriesONC} that observational and physical effects give rise to the observed luminosity dispersion. Because of the observed colour spread in IC~348 \cite{Bell} argues for an age of 6\,Myr for IC~348 while \citep{luhmanMembers} adopted a median age of 2\,Myr which we use in this work.

The stellar content of IC~348 was researched with the photometric studies by \cite{herbig} and \cite{luhmanIMF}. With the help of the proper motion measurements from \cite{scholz} and additional photometric and spectroscopic data \cite{luhmanMembers} compiled a membership catalogue of IC~348, including spectral classification for most of its members. This catalogue has recently been extended by \cite{luhmannew} to include 478 members. The membership criterion was based on proper motion, position on the colour-magnitude diagram, and spectral features and classification.

Using CCD detectors, photometric time-series of different durations and with different fields of view (FoV), but all in the $I$ band, have been obtained and published. The first time-series of its kind of IC~348 was published by \cite{herbst} with a baseline time-scale of four months. The FoV was $10.2\arcmin\,\times\,10.2\arcmin$ and the authors found 19 periodic variable stars near the centre of the cluster. With more data from the same survey, \cite{CohenHerbst} were able to detect 28 periodic variables. Finally, with a baseline of seven years, \cite{nordhagen} found twelve additional periodic stars in the same FoV. An independent study with deeper photometry was published by \cite{littlefair} and found 32 new periodic variables in IC~348. The most extensive study so far has been conducted by \cite{cieza}. The authors used a wider FoV ($46.2\arcmin\,\times\,46.2\arcmin$) and were able to discover 75 additional periodic variables. Moreover, they combined and analysed all previous studies and counted a total of 106 periodic stars among the members of IC~348 when applying the membership of \cite{luhmannew}.

The evolution of angular momentum in young stars is a topic of recent research (e.g. \citealt{Tanveer}) and the distribution of rotation periods is a stepping stone to its understanding. Since the discovery of a bimodal period distribution for stars with $M > 0.25 M_{\sun}$ in the 1\,Myr old Orion Nebula Cluster (ONC) by \cite{ONCherbst} other open clusters are compared to this young cluster. IC~348 is slightly older and the previous time-series studies came to the conclusion that the period distributions of both clusters look alike. The distribution consists of fast rotators with periods of $\sim 2\,\mathrm{d}$ and slow rotators with periods of $\sim 8\,\mathrm{d}$. With more rotational periods for the members of IC~348 we can investigate the distribution in more detail.

In this study, carried out within the YETI network we used an even wider FoV ($52.8\arcmin\,\times\,52.8\arcmin$) than any previous studies. With this FoV we can find variabilities in IC~348 and its vicinity. Moreover, we used several telescopes located all over the world to achieve a better phase coverage of our time-series.

The paper starts with an overview of our observations and the data reduction workflow (\autoref{sec:obs}). Thereafter, we present the results (\autoref{sec:res}), starting with variables in IC~348, followed by further results on field stars. In the discussion (\autoref{sec:dis}) we compare our results with previous studies.

\section[]{Observations and data reduction}
\label{sec:obs}
\subsection{Observations}

\begin{figure*}
	\includegraphics[width=\textwidth]{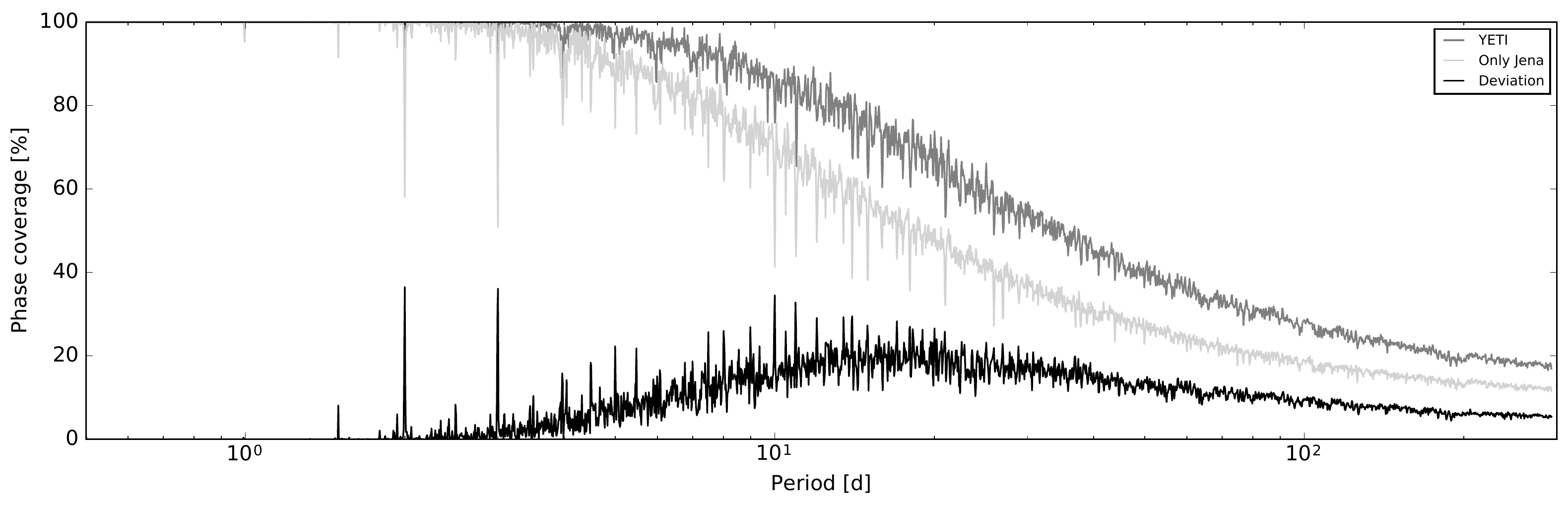}
	\caption{Phase coverage of the data for a single star. The dark grey line includes all YETI observations while the light grey line gives the phase coverage for the observations from the University Observatory Jena. In black the difference of both curves is shown for easier interpretation. The better phase-coverage for periods of the multiple of one day is clearly visible (spikes), as well as the big advantage for periods in the range of 10\,d to 25\,d.}
	\label{fig:phasecover}
\end{figure*}

IC~348 was observed between 2012 August 22 and 2015 January 18 in 125 telescopic nights over three seasons.
Observations were carried out by nine telescopes through the YETI network, with a distributed longitudinal coverage (\autoref{tab:sites}). The University Observatory Jena alone contributed 88 nights.
Details on all observations can be found in \autoref{tab:observations}. To achieve a good phase coverage three week-long YETI campaigns were performed in which IC~348 was observed from the participating observatories on every clear night. Therefore continuous coverage of IC~348 could be achieved at some nights through combined observations. Overall, the distributed observations resulted in a better phase-coverage of the data. In \autoref{fig:phasecover} we compare the phase-coverage of YETI data to Jena data alone and the advantage of YETI is clearly visible. For easier interpretation of \autoref{fig:phasecover} we show the deviation of the curves in addition to the data. The combined light curves, compiled from all telescopes, have better phase-coverage in the range of 10\,d to 25\,d (25 percentage points) and for periods of the multiple of one day (up to 35 percentage points). The better phase-coverage for multiple periods of one day reduces the 1\,d alias period significantly when searching for periods.

\begin{table*}
	\caption{Overview of the YETI telescopes used for this work. FoV stands for field of view. In Star\'a Lesn\'a two 0.6\,m telescopes have been used with different detectors. The table is sorted by the number of obtained frames.}
	\label{tab:sites}
	\begin{tabular}{l l r r c r l r}
		\hline
		Observatory & Telescope & Longitude & Latitude & FoV & \#\,Frames & CCD Detector & Pixel scale\\
			& & [\degr] & [\degr] & [\arcmin] & & & [\arcsec / px] \\
		\hline
		Jena$^a$ & 0.6\,m Schmidt & 11.48416\,E & 50.92888\,N  & 52.8\,$\times$\,52.8	& 10\,808 & e2v 42-10 & 1.547\\
		Jena$^b$ & 0.25\, Cassegrain & 11.48416\,E & 50.92888\,N & 21.0\,$\times$\,20.4 & 2390 & e2v 47-10 & 1.193\\
		Star\'a Lesn\'a & 0.6\,m Cassegrain& 20.29081\,E & 49.15207\,N & 14\,$\times$\,14 & 1\,957 & FLI ML3048 & 0.410\\
		& & & & 16.9\,$\times$\,16.9 & & MI G4-9000 & 0.333\\
		pt5m, La Palma$^c$ & 0.5\,m mod. Dall-Kirkham & 17.88188\,W & 28.76058\,N & 10.2\,$\times$\,6.9 & 1\,050& KAF-3200ME & 0.280\\
		Lulin & 1\,m Cassegrain & 120.87297\,E & 23.46908\,N & 11.5\,$\times$\,11.2 & 614& e2v 36-40 & 0.515\\
		Tenagra & 0.8\,m Ritchey-Chrétien & 110.98475\,W & 31.55571\,N & 14.8\,$\times$\,14.8 & 584& & 0.867\\
		Saitama & 0.55\,m Ritchey-Chrétien & 139.60561\,E & 35.86225\,N & 12.8\,$\times$\,12.4 & 218& ML4710-1 & 0.727\\
		Swarthmore& 0.6\,m Ritchey-Chrétien & 75.35605\,W & 39.90702\,N & 26\,$\times$\,26 & 195& Apogee U16M  & 0.381\\
		Llano del Hato & 1\,m Cassegrain-Coudé & 70.87094\,W & 8.78794\,N & 18.5\,$\times$\,18.5 & 30& e2v 42-40 & 0.542\\
		\hline
	\end{tabular}
	\flushleft{References: $^a$\cite{MugrauerSTK}, $^b$\cite{MugrauerCTKII}, $^c$\cite{pt5m}}
\end{table*}

\begin{table}
	\caption{Observational seasons and number of nights with successful observations.}
	\label{tab:observations}
	\begin{tabular}{rlr}
		\hline
		Season & Date & Number of nights\\
		\hline
		1 & 2012 Aug. 8 -- 2013 Mar. 15 & 57\\ % (36,-,-,1,2,3,15,- )\\
		2 & 2013 Aug. 1 -- 2014 Feb. 4 & 41\\ % (30,6,-,1,4,-,-,-)\\
		3 & 2014 Sep. 16 -- 2015 Jan. 18 & 27\\ % (22,-,3,-,-,-,-,2)\\
		\hline
	\end{tabular}
\end{table}

In Jena the open cluster was observed with the 0.6\,m Schmidt telescope. We used the Schmidt-Teleskop-Kamera (STK) \citep{MugrauerSTK} with its Bessel $R$ filter and exposed for 50\,s. The exposure times at other telescopes differed because of different apertures, smaller FoV, and other detectors. Additional $BVI$ images were acquired on some nights at the University Observatory Jena. Further observations were carried out with the 0.25\,m Cassegrain telescope equipped with the Cassegrain-Teleskop-Kamera II (CTK-II) \citep{MugrauerCTKII} at the University Observatory Jena. For those observations we used Bessel $V$ and $I$ filters and exposed for 180\,s.

In addition to the Jena telescopes we observed with the pt5m telescope on La Palma \citep{pt5m}, the 1\,m Cassegrain at the LOT observatory in Lulin, the 1\,m Cassegrain-Coudé in Llano del Hato, the two 0.6\,m Cassegrain in Stará Lesná, and the Ritchey-Chrétien telescopes of Saitama University, Swarthmore College, and Tenagra observatory. The locations, diameters, and detectors of all telescopes used can be found in \autoref{tab:sites}.

Most data were gathered with the STK operated at the Jena 0.6\,m Schmidt telescope, therefore the stars which were analysed are all in the FoV of this telescope. Additionally, the STK exhibits the widest FoV of all instruments used. Some telescopes obtained images with deeper photometry. We decided not to use the additional fainter stars from those frames. The sparse coverage of the light curve would not have given insight to the time-scale we are interested in.

The STK is equipped with a 2048\,px\,$\times$\,2048\,px e2v CCD detector with a FoV of $52.8'\times52.8'$ \citep{MugrauerSTK}. The centre of IC~348 was positioned slightly off the centre of the detector to north-east to allow simultaneous observations with the CTK-II (centred at $\alpha = 3^\mathrm{h}\,45^\mathrm{m}\,20^\mathrm{s}$,~$\delta = +32\degr\,4\arcmin\,50\arcsec$). From this FoV the 1001 brightest stars have been analysed independent of their membership status down to a limiting magnitude of $R=(18.7\pm0.3)\,\mathrm{mag}$. Within our sample 137 stars are members of IC~348 according to \cite{luhmannew}. Further stars are either unidentified members, background or foreground stars.

All other telescopes used for this observations have a smaller FoV. Therefore, they included a tighter area around IC~348 or covered the whole region with various pointings, which implies that not all stars of our selection have been observed by all telescopes.

\begin{figure*}
	\includegraphics[width=\textwidth]{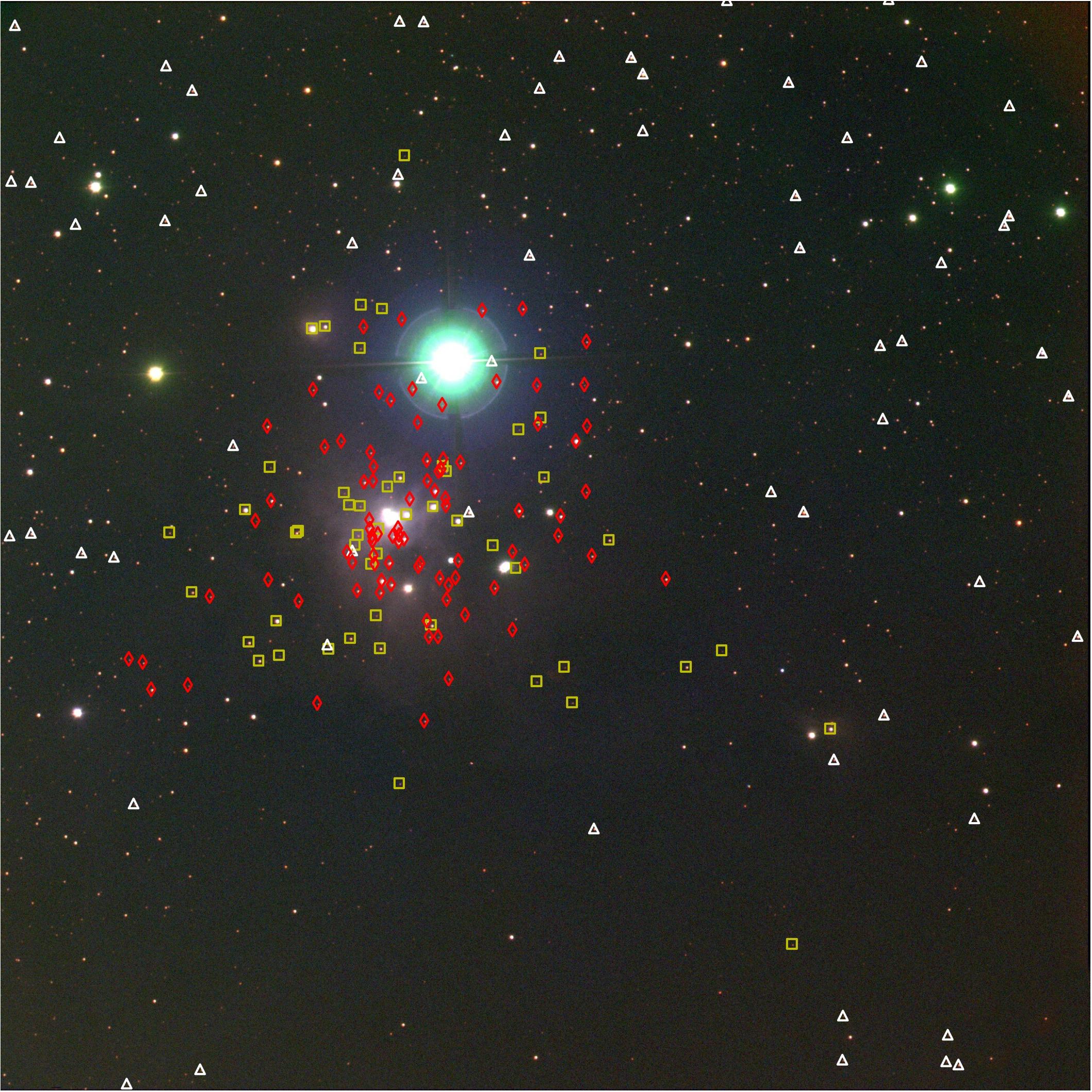}
	\caption{$B$$V$$R$ image of IC~348 as observed from Jena with STK on 2014 December 8 with an integration time of 50\,s in each band. The full STK field of view with a size of $52.8\arcmin \times 52.8\arcmin$ is shown. North is up and East to the left. The open cluster IC~348 is located around and within the reflection nebula south of the bright star Atik (o Per). Periodically variable members of IC~348 are marked with a (red) diamond, while all other members detected in this image are marked with a (yellow) square. Non-members for which we found a period are marked with a (white) triangle.}
	\label{fig:bvr}
\end{figure*}

The FoV of the STK is shown in \autoref{fig:bvr} as a composite image of $B$, $V$, and $R$ observations. It is dominated by the bright star Atik (o~Per). The open cluster IC~348 lies south of it and is surrounded by a reflection nebula. In this frame we marked our observed stars according to their properties.

\subsection{Data reduction and photometry}
For every night, in addition to the science frames, darks and flats were obtained. Whenever possible we acquired sky flats, otherwise the nightly dome flats were used. Some telescopes included the bias in an overscan region while others produced dedicated bias frames. From those images a standard reduction with dark, bias, and flat field correction was performed with \textsc{iraf}.

For every telescopic pointing the data were extracted separately from the reduced images with aperture photometry followed by differential photometry. For the differential photometry we used an implementation of the algorithm presented by \cite{broeg}, based on the \textsc{iraf} task \textsc{phot} as described in detail in \cite{errmann}. This algorithm calculates an artificial star for comparison from all stars in the FoV, weighted according to the standard deviation of the differential light curves. From this procedure we obtained a light curve for each pointing on each night.

\begin{figure}
	\includegraphics[width=\columnwidth]{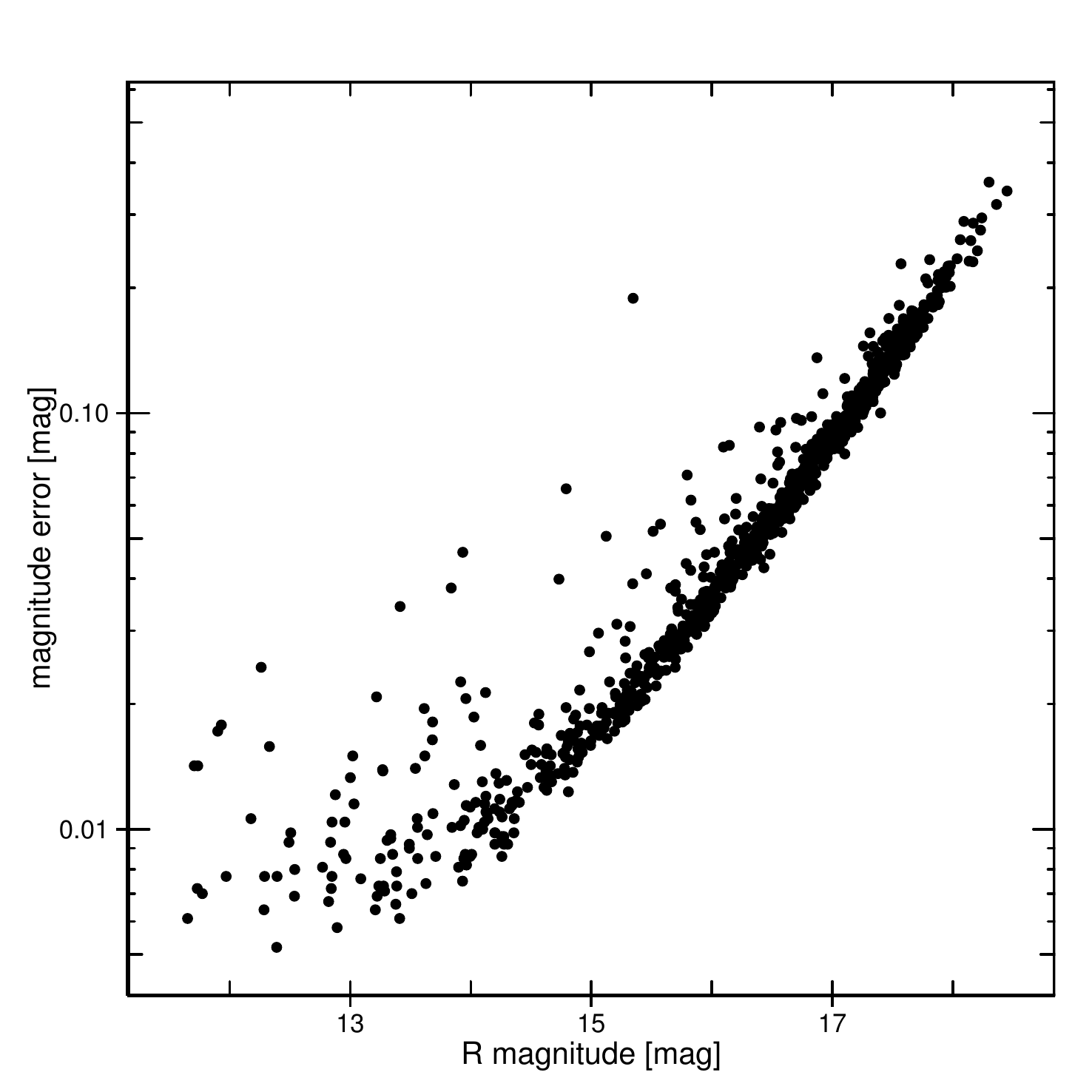}
	\caption{Photometric precision for the night of 2014 December 8 for the observations at the University Observatory Jena. This figure includes all 1001 selected stars from the FoV. In that night we also gathered $B$$V$$I$ photometry of IC~348 with the STK.}
	\label{fig:photoprec}
\end{figure}

In \autoref{fig:photoprec} the mean photometric precision of one night is shown. The winter night of 2014 December 8 is exemplary and represents the normal conditions achieved at the University Observatory Jena. E.g. for a 16\,mag star a photometric precision of 0.03\,mag is reached.

After combining the data for each telescope, by adjusting the flux level from night to night, we had to collate all observations of each star obtained with different telescopes. For this step we first searched for periodic variations in the light curves from the Jena observations using the algorithms presented in \autoref{sec:alg}. Thereafter, we used the periods to produce phase-folded light curves. Because of the tighter sampling in the phase domain the data from other telescopes can easily be incorporated into the light curve. This was done by adjusting the flux of the observations to the flux of the data with similar phases. The very good phase-coverage of the Jena data was mandatory for this method.

With only a few nights of multi-band photometry we decided to anchor our photometry to \cite{trullols}. We used the night of 2014 December 8 for the transformation and all magnitudes used in the current work are based on this system. 

\subsection{Algorithms used to find periodic variabilities}
\label{sec:alg}

The main goal of this work was to find new periodic photometric variables in IC~348 and to confirm and improve ephemerides of already known variable stars in this cluster. Therefore we applied three different algorithms to detect periodicities in the data. All algorithms have in common that they use a grid of fixed periods as their input. We employed the same grid for all algorithms to gain comparable results.

This period grid had a range from $P_\mathrm{min} = 0.04\,\mathrm{d}$ ($\sim 1\,\mathrm{h}$) to $P_\mathrm{max} = T/3=293\,\mathrm{d}$ (where $T$ is our observational timebase). We used such a wide range because at least one long-periodic eclipse is known in IC~348 \citep{nordUnusal} and more might be discovered. We chose our lower limit to include short-term variability but not stellar pulsations. The upper limit was guided by the decision to include at least three cycles. To save computation time we used an exponentially-spaced grid with $n \sim38\,000$ points. The resolution of the grid is for the lower limit 1\,s, for a period of 1\,d 20\,s and for the upper limit of 293\,d 1.6\,h.

The first algorithm we used is the widely applied generalized Lomb-Scargle (GLS) periodogram by \cite{zechmeister}. From the periodogram we obtained the spectral power density as a measurement of the certainty of the periodic variations. Although the algorithm is fast and convenient for finding periods it has some drawbacks. It favours signals with sinusoidal shape which can lead to incorrect best-fitting values for non-sinusoidal-shaped light curves like transits, occultations or eclipsing binaries. 

As a second algorithm the minimal string-length algorithm by \cite{dworetsky} was used. Unlike the GLS a string-length algorithm can find periodic variabilities of all shapes with the same sensitivity. In this algorithm the sum of the distances of succeeding points in the phase-folded representation is measured. Because of lower phase-coverage for long periods this method is biased towards longer periods. To correct this effect a sum of a second-order polynomial and an exponential decay was fitted to the output of the algorithm. From the normalized and detrended results we were able to find the best period.

Our third algorithm was the Gregory-Loredo Bayesian signal detection as presented in  \cite{gregory} and \cite{gregory99}. This method uses different step functions to calculate the likelihood for a given period based on Bayesian statistics. It is unbiased towards shape and sampling of the light curve because the step function can adopt arbitrary shapes. We used the assumption of independent Gaussian errors and directly applied the formulation given by \cite{gregory99}.

Each algorithm returned a best period leading to three different values after one run of the algorithms. To find the best period automatically we applied an additional program. Therein we set as an initial selection criterion that two of the three algorithms find the same period within an error range of ten per cent. Afterwards an additional run of the period search was conducted within that ten per cent range of the best period. Now a match within one per cent was required and we used a tighter spacing of the grid. With this criterion we were able to use the maxima of the periodograms independently of their power density and were able to detect variabilities with low signal-to-noise ratio. To exclude false-detection all phase-folded light curves were examined manually and non-periodics were removed.

\section[]{Results}
\label{sec:res}

In this section we first present the results for the member stars of IC~348 and later findings for non-members. The membership is according to \cite{luhmannew}.

If a star has a commonly used name we use this name to identify it in this section and give other identifiers as a footnote. Otherwise we use our internal numbering. The abbreviations used in this section are as following: LRL refers to \cite{luhmanIMF} (and subsequent publications), HMW to \cite{herbst}, CB to \cite{cieza}, and FKM to this work. When no distinct identifier is known we give the 2MASS name but continue using our identifier.

\subsection{Periodic variables in IC 348}

With the above mentioned method we were able to identify 87 photometric periodic stars in IC~348. Of those stars 33 have not been reported as periodic before. Including all previous studies (overview of \citealt{cieza}) the total number of periodic variables in IC~348 is now 139 out of 478 members. The reasons for non-detections of previously known periodic variables are discussed in \autoref{sec:compare}. Most of the stars are rotating young stars that show spot induced variability, although some periodic variabilities are due to occultations of proto-planetary discs or accretion. The results are summarized in \autoref{tab:allperiods} and the phase-folded light curves are shown in \autoref{fig:lcs}. In the following paragraphs we will present only some notable variables.

\subsubsection{V695 Persei}

\begin{figure*}
	\includegraphics[width=\textwidth]{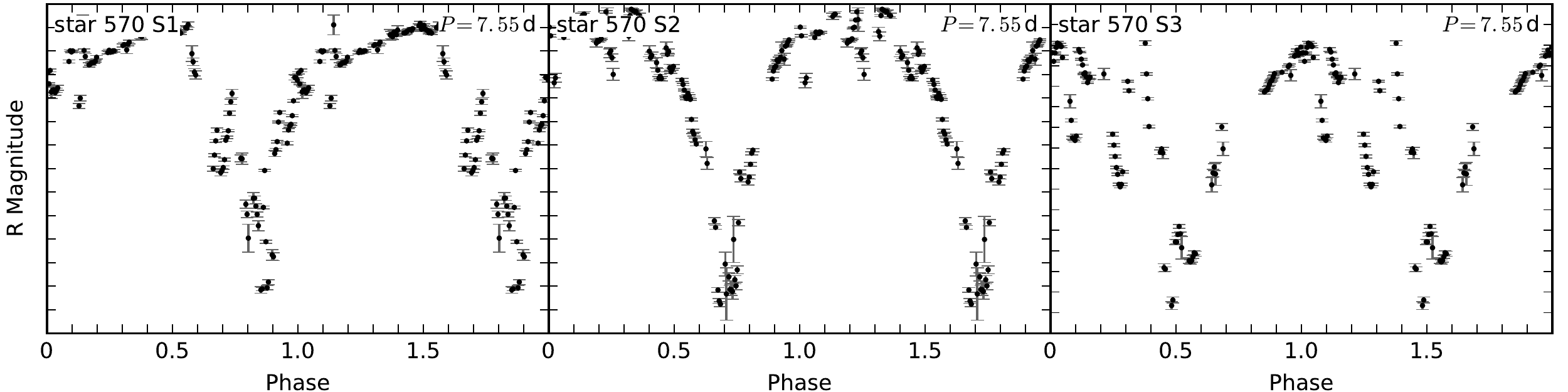}
	\caption{Phase-folded light curves of the AA Tauri-like star V695 Per for all three observing seasons. While the first two seasons are phased adequately the observations from the third season show some variability in the phase. The data has been binned and the spacing is 0.1\,mag for all tick marks on the magnitude axis.}
	\label{fig:V695}
\end{figure*}

V695 Per\footnote{FKM 570, LRL 99, HMW 73, CB 49} has one of the largest peak-to-peak amplitudes in our observations with 1.2\,mag. The phase-folded light curve (\autoref{fig:V695}) resembles a typical occultation which was explained by \cite{V695} as an AA Tauri-like system. The proto-planetary disc follows a Keplerian motion with a period of $(7.55\pm0.08)\,\mathrm{d}$. For the third season we observed a change in the light curve. Some data points seem to be outlier to the previously observed shape of the light curve. This is shown in the third panel of \autoref{fig:V695}. The outliers might be explained with a reconfiguration of the proto-planetary disc or the magnetic field. As a result the disc could be warped in a different way which leads to a change of the phase of the occultation. Alternatively an additional occultation might take place and the two effects overlap. Further monitoring of this star can help to constrain the reasons.

\subsubsection{V718 Persei}

\begin{figure}
	\includegraphics[width=\columnwidth]{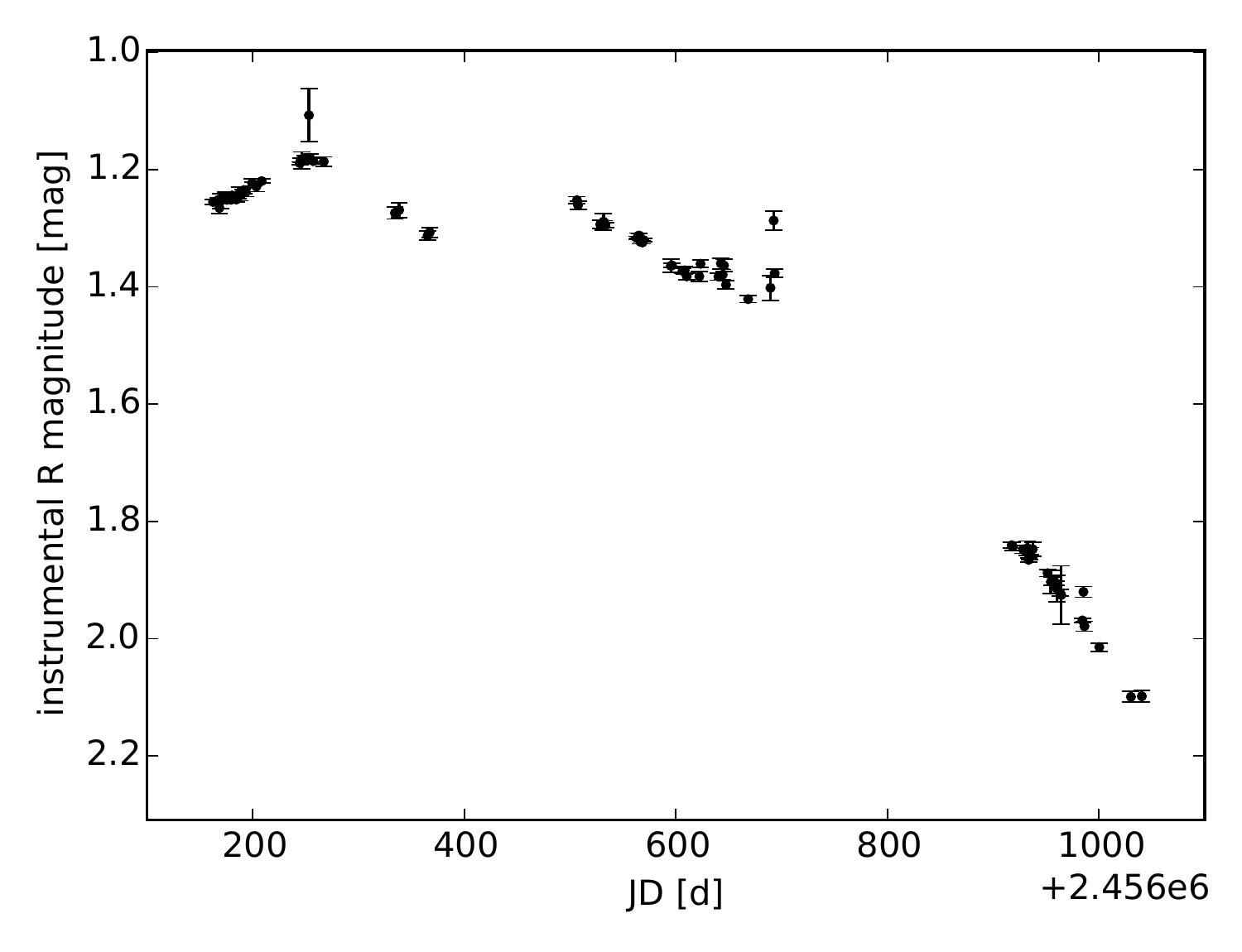}
	\caption{The full light curve of V718 Per shows the strong decrease of brightness due to the long-periodic occultation of the star. The data have been binned to one data point per night.}
	\label{fig:V718}
\end{figure}

The unusual photometric periodicity of the star V718 Per\footnote{FKM 123, HMW 15, LRL 35} was discovered by \cite{CohenUnusual} and analysed in depth thereafter \citep{nordUnusal, Grinin}. It is periodically occulted by a part of its proto-planetary disc. The occultation has a period of $(4.7\pm 0.1)\,\mathrm{yr}$ \citep{nordUnusal} while the eclipse lasts for 3.5\,yr. In our data, spanning 2.4\,yr, we were able to see the decrease of brightness of $0.9$\,mag (\autoref{fig:V718}). This observation shows that V178 Per is a rather stable system. IC~348 was monitored from Van Vleck Observatory from 1991 \citep{nordUnusal}: the system was therefore observed in a stable configuration for 24 years now.

\subsubsection{A close stellar companion to LRL 47}

\begin{figure}
	\includegraphics[width=\columnwidth]{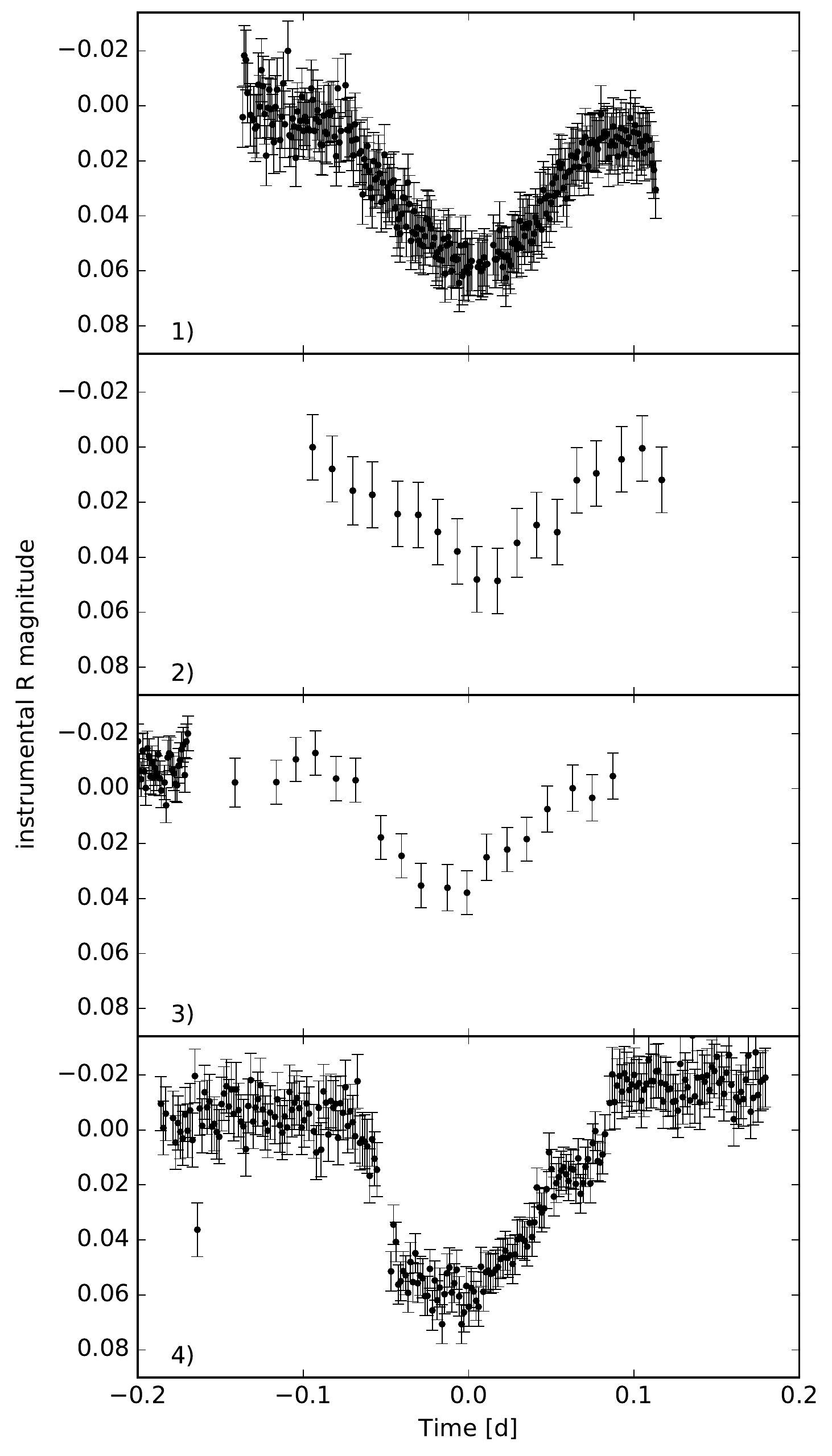}
	\caption{The light curves of the four observed transits of LRL~47 (star 71). Each is centred on the transit mid-point (see \autoref{tab:transitdata}) and scaled the same for better comparison. The Tenagra telescope (2 and 3) has observed with a lower sampling than the other two telescopes (Jena/STK panel 1 and Lulin panel 4) therefore the data are sparse. In panel 3 the advantage of the YETI network is visible. The observations from two telescopes are included in this light curve. All information on the transits are listed in \autoref{tab:transitdata}, including depth, duration, telescope, and time.}
	\label{fig:transits}
\end{figure}

As well as the modulation of the flux due to starspots with a period of $(4.91\pm0.05)\,\mathrm{d}$, the light curve of LRL 47\footnote{FKM 71} showed an additional feature. In the three observing seasons four short dips were found (see \autoref{fig:transits}). All of them show the typical V-shape of a grazing transit.

LRL 47 is known to be active in X-ray \citep{preibisch, PZ01, PZ02, Stelzer, FlahertyXray} like a lot of young stars are \citep{NeuhauserSci}. \cite{lada} found no evidence of a circumstellar disc around it. \cite{dahm} and \cite{currie} later confirmed this result. The non-existence of a a disc is not unusual for stars in IC~348. \cite{lada} found a disc fraction of $(50\pm6)$ per cent and \cite{cieza07} state that discs might disappear as soon as 1\,Myr after the stellar formation, building planets within this time. This star might have had a circumstellar disc which has already vanished. In addition to the non-existence of a disc, \cite{duchene} were able to exclude a stellar companion to this star with a detection limit of 3\,mag at 0.5\arcsec (i.e. 158\,AU).

The transits were observed at Jena, Tenagra, and Lulin observatories. Hence, we have three independent observations, an observational effect can therefore be excluded. These observations show again the power of the YETI network. With only a single telescope we would have missed most of the events. From the University Observatory Jena only the first event was visible and was observed. Unfortunately the $V$ and $I$ band observations carried out with the CTK-II mounted to the second telescope at the University Observatory Jena are not accurate enough to gain additional colour information of the transit. 

The properties for the four transits are given in \autoref{tab:transitdata}. The transit mid-points has been obtained with \textsc{jktebop} \citep{southworth, Etzel, PopperEtzel}. Also with \textsc{jktebop} we determined a period of $(5.123874\pm0.000063)\,\mathrm{d}$ from those four transits. This period is close to the rotational period of $(4.91\pm0.05)\,\mathrm{d}$ but is feasible for a close companion.

\begin{table}
	\caption{Parameters of the observed transits of LRL~47. MJD$_\mathrm{mid}$ gives the midpoint of the transit as modified JD ($\mathrm{MJD} = \mathrm{JD} - 2\,456\,000\,\mathrm{d}$). The epoch is the orbital cycle number relative to the first observation.}
	\label{tab:transitdata}
	\begin{tabular}{r r c r l }
		\hline
		Number & Depth & MJD$_\mathrm{mid}$ & Epoch & Telescope\\
		& [mag] & [d] & &\\
		\hline
		1 & 0.05& $188.5105\pm 0.0016$& 0 & Jena\\
		2 & 0.05& $203.8714\pm 0.0149$& 3 & Tenagra\\
		3 & 0.04& $244.8718\pm 0.0097$& 11 & Tenagra\\
		4 & 0.07& $962.2088\pm 0.0030$& 151 & Lulin\\
		\hline
	\end{tabular}
\end{table}

According to \cite{luhmanMembers} LRL 47 is a $\mathrm{K}0 \pm 2$ star. From the isochrones of \cite{siess} we can find for the spectral type K0 a mass of $M_\mathrm{P} = 2.7\,\mathrm{M}_{\sun}$ and a radius of $R_\mathrm{P} = 4.14\,\mathrm{R}_{\sun}$ at the age of 2\,Myr. This age is not necessarily the true age of this star. \cite{luhmanIMF} gave an age-spread from 0.5 to 10\,Myr for IC~348 which is under debate (see \autoref{sec:intro}). Therefore the primary's radius can be overestimated by a factor of two. Nevertheless, we continue with the median age of 2\,Myr to find an estimation of the companion's properties.

Using \textsc{jktebop} again for fitting the data did not lead to a satisfactory result. The reason for that is the grazing nature of the transit and our lack of additional information on the system. To get an idea of what kind the transiting object is, we assumed a Keplerian orbit and a non-grazing transit. With these assumptions we can calculate lower boundaries for the companion's radius and mass.

First, we calculated the inclination and radius from our supposed values for the primary and the measured parameters of the transit. Thereafter we derived the mass for the companion from the low-mass isochrones of \cite{BaraffeNew}. After an iterative process we concluded on an inclination of $i \lesssim 75.4\degr$, a radius of $R_\mathrm{C} \gtrsim 0.87\,\mathrm{R}_{\sun}$, and a mass of $M_\mathrm{C} \gtrsim 0.09\,\mathrm{M}_{\sun}$. Here we give no uncertainties because they would strongly underestimate the true errors. As said above those values are only lower (upper) boundaries for the mass and radius (and inclination). Due to the grazing of the transit we cannot be sure about the true radii ratio and therefore the derived values. From those lower boundaries we can conclude that this star might have a close stellar companion which might be near the hydrogen burning limit.

The four light curves show two slightly different shapes (\autoref{fig:transits}). The first two occurrences of the transit have a symmetric V-shape. In contrast, the later two have a steep decrease of flux at the beginning and a slower increase after the minimum. Even though two different shapes are observed we are not looking at a secondary and a primary transit. Except the first transit all have an odd orbital cycle number which rules out the double period. A secondary transit is not visible in the data of phase 0.5. This constrains the mass ratio to $\frac{M_\mathrm{C}}{M_\mathrm{P}} \ll 1$. The different shapes of the transit can be explained with active regions and starspots on the stellar surface. 

Independent of the mass, this short period means that the companion has a small semi-major axis in the range of $0.08\,\mathrm{AU}$ to $0.1\,\mathrm{AU}$. Therefore it is close to the primary and could not have been detected by direct imaging even using adaptive optics on a 4\,m class telescope \citep{duchene}.

To constrain the mass of the companion candidate radial velocity measurements of the primary with $R=13.5\,\mathrm{mag}$ are necessary.

\subsubsection{Period distribution}

The distribution of the rotation periods has been an interest to all previous time-series studies of IC~348. It has always been compared to the slightly younger ONC for which \cite{herbstONC2000} discovered a bimodal period distribution. The bimodal distribution is present for more massive stars which was defined by \cite{herbstONC2000} as $M > 0.25 M_{\sun}$ or SpT earlier than M2 \citep{CohenHerbst} for young stars.

The comparison always gave the impression that the period distribution is similar to the period distribution of the ONC. Hence, it is bimodal with peaks for periods of about 2\,d and 8\,d. Despite the similarities a Kolmogorov-Smirnov test could not confirm that the two distributions (ONC and IC~348) are the same \citep{CohenHerbst, littlefair}. Furthermore, \cite{littlefair} used the Hartigan dip test \citep{hartigan, hartigan2} and found that the bimodality in IC~348 is not statistically significant. Using the \textsc{r} \citep{R} implementation of the dip test \citep{diptest} we find that the extended set of rotation periods now available has no statistically significant bimodality.

\begin{figure}
	\includegraphics[width=\columnwidth]{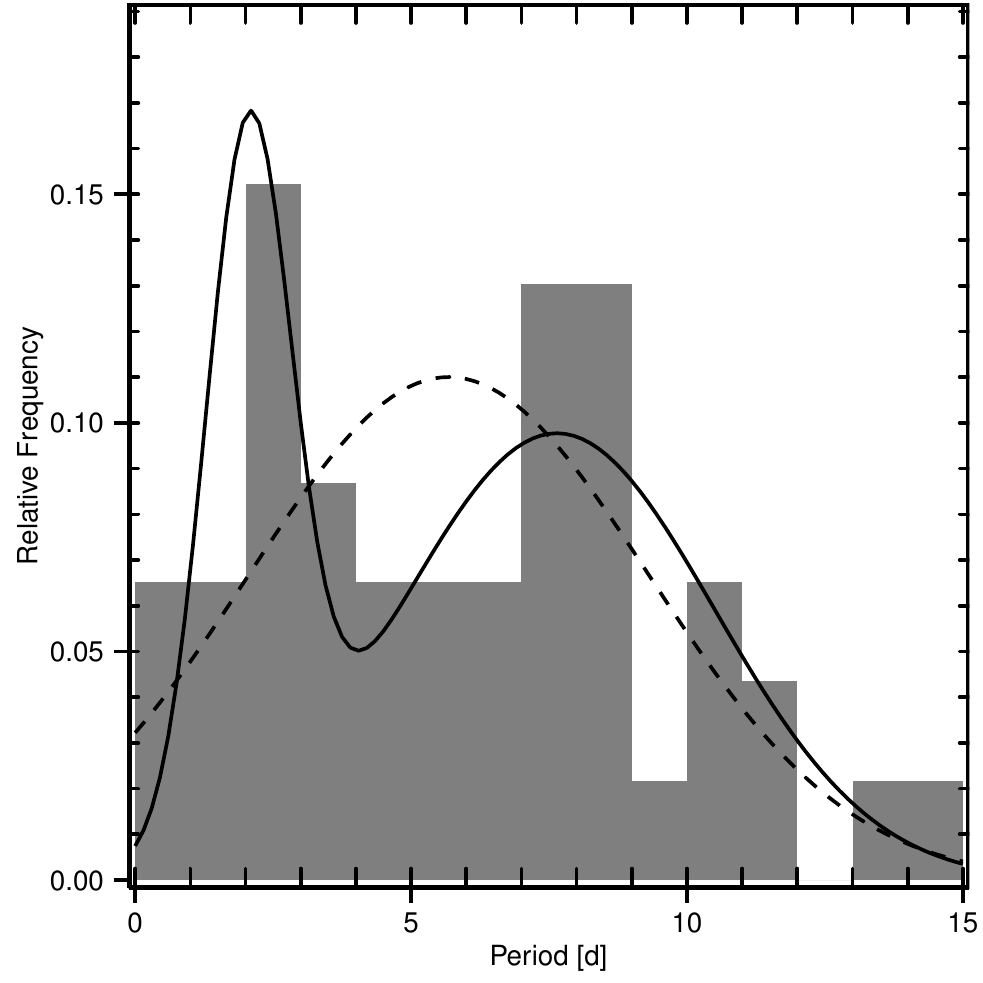}
	\caption{Period distribution of the rotation periods for high-mass stars in IC~348 (grey bars). Overlaid are a simple Gaussian model (dashed line) and a bimodal mixture model of two different Gaussians (solid line).}
	\label{fig:bimod}
\end{figure}

\begin{figure}
  \includegraphics[width=\columnwidth]{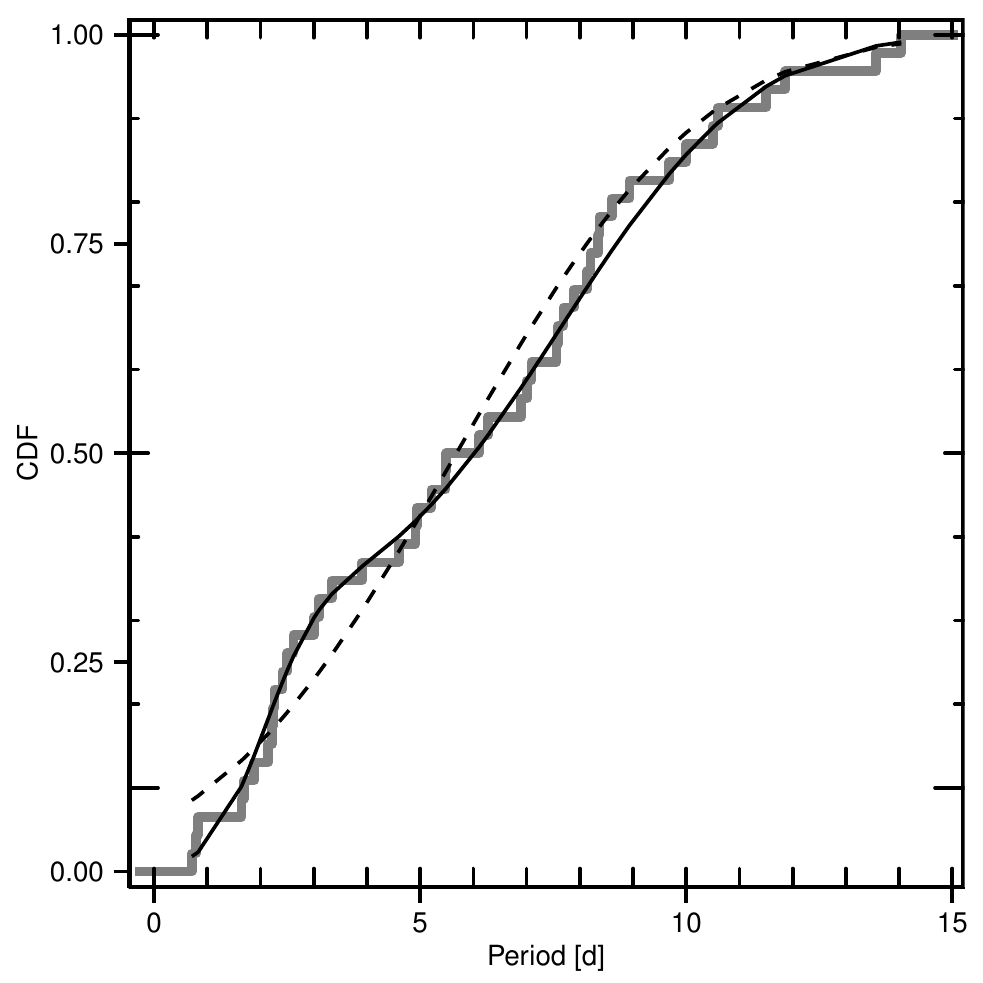}
  \caption{Cumulative distribution function (CDF) of our data. The mixture model is fitting very well. The colours and line types are the same as in \autoref{fig:bimod}.}
  \label{fig:bimodCDF}
\end{figure}

In addition to the dip test we applied two Gaussian models (\autoref{fig:bimod}) to the data. At first we used the \textsc{mass} package in \textsc{r} \citep{RMASS} to fit an unimodal Gaussian to the data. The second model is a mixture model of two different Gaussians. To find the parameters we applied the \textsc{mixtools} package for \textsc{r} \citep{mixtools}. Both models were compared to the empirical cumulative distribution function (eCDF) of the data (\autoref{fig:bimodCDF}). An F-test yields a $p$-value of $1.1\cdot10^{-9}$, which is the probability that the null hypothesis is true. In this case the null hypothesis is the unimodal distribution. Including all data from \cite{cieza} the F-test yields an even smaller $p$-value.

We conclude that a bimodal Gaussian distribution fits more likely the data than a unimodal Gaussian distribution. Since the dip test, in contrast, favours the unimodal model the bimodality of the rotation period distribution in IC~348 is still an open issue.

\begin{figure}
	\includegraphics[width=\columnwidth]{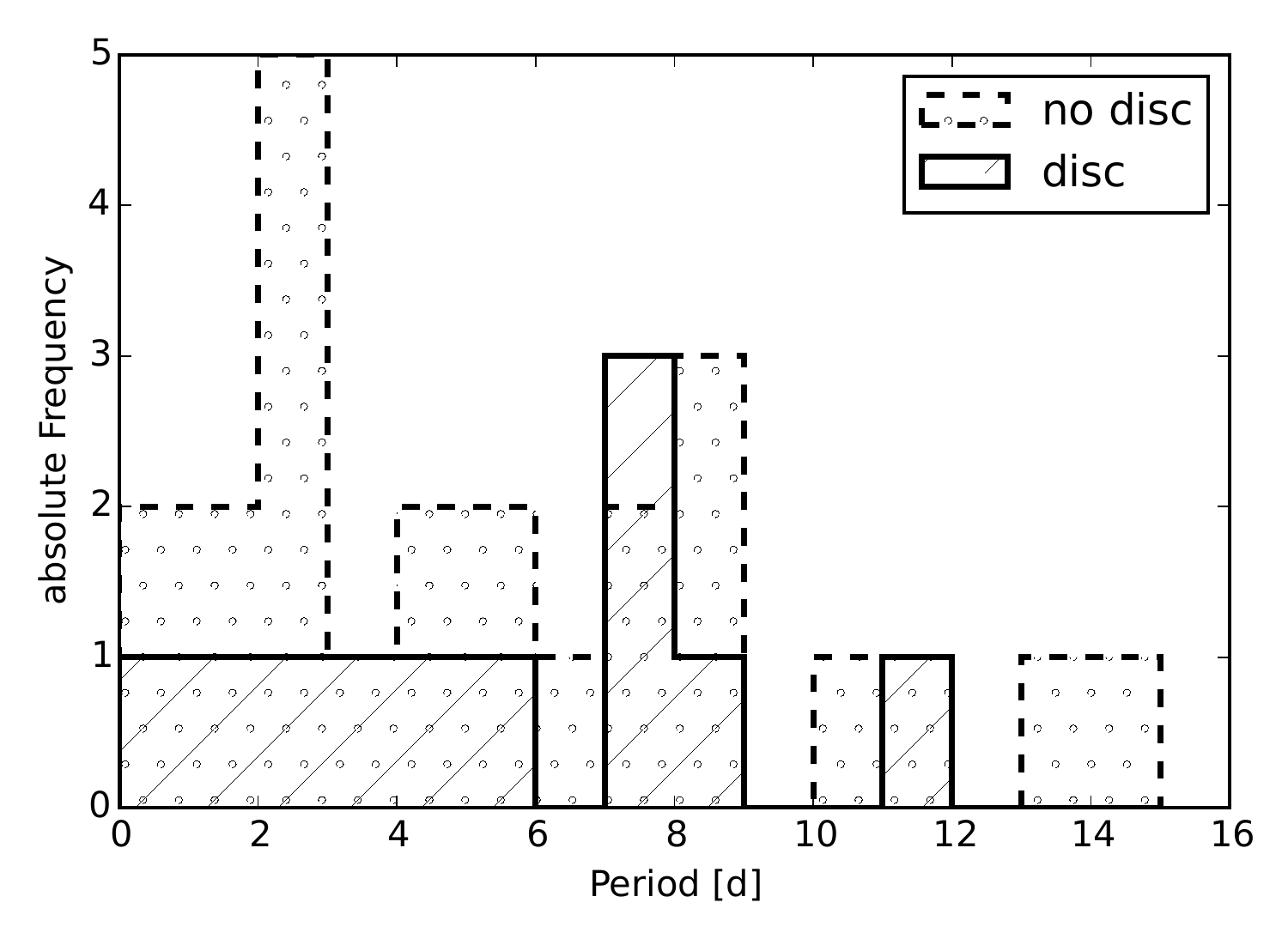}
	\caption{The period distribution of stars with $M > 0.25 M_{\sun}$ separated by the existence of discs. Few discs are observed around those more massive stars in the cluster and we cannot make a statement about influence of the disc on the stellar rotation.}
	\label{fig:discPeriods}
\end{figure}

Separating the rotation periods for stars with and without discs using the data from \cite{lada} is also inconclusive. The low numbers of stars with discs shows a peak for the slow rotators (\autoref{fig:discPeriods}) but with only two more stars than in the other bins. Among the stars without an observed disc indicator more fast rotators can be found. Nevertheless, slow rotators are among those stars, too. For the histogram in \autoref{fig:discPeriods} we used only stars which are in the group of $M > 0.25 M_{\sun}$. \cite{cieza} also found no evidence that the slow rotors preferably have discs. Even with the additional periods the statistics are still inconclusive.  

\begin{figure}
	\includegraphics[width=\columnwidth]{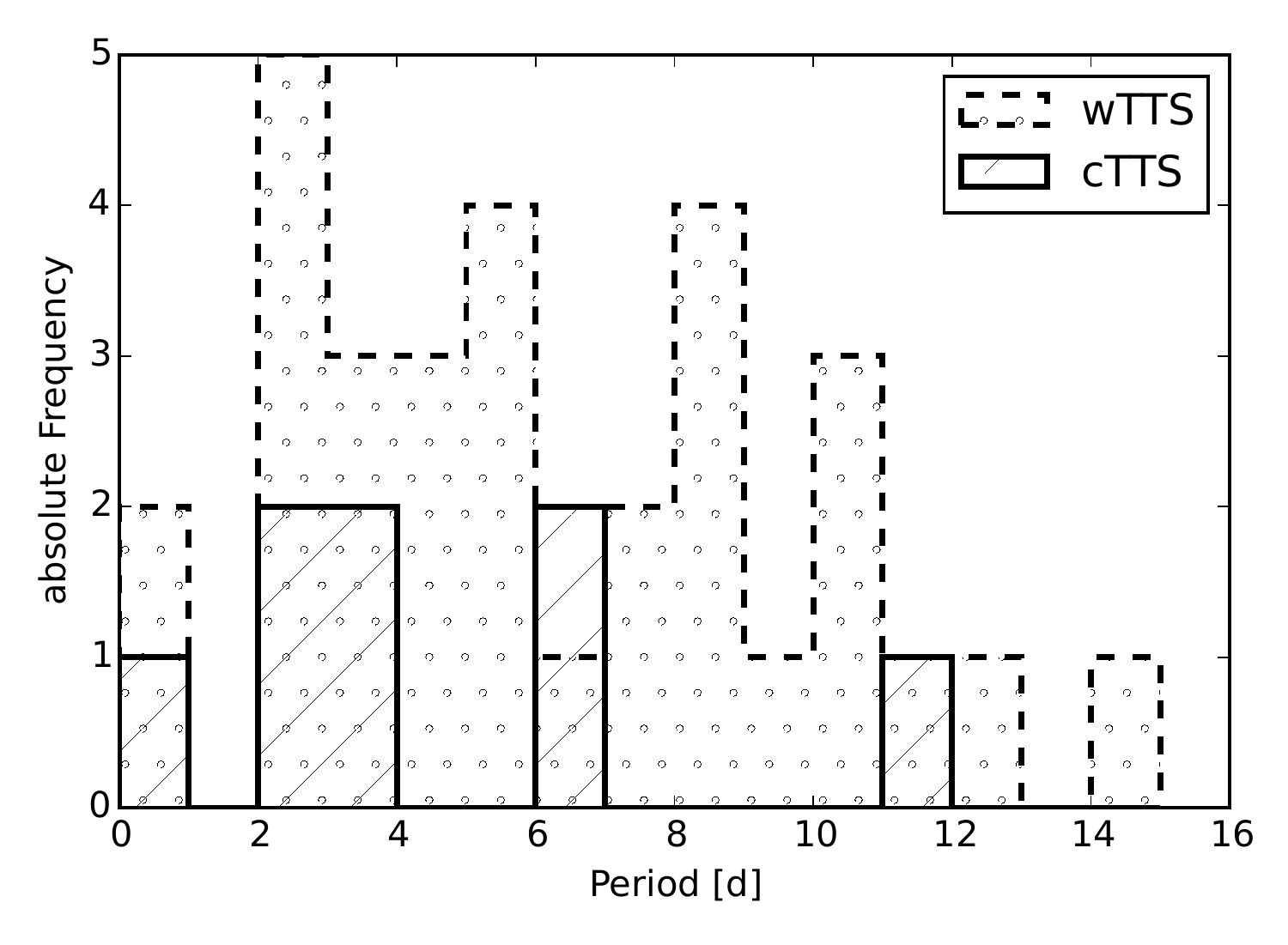}
	\caption{The period distribution of T-Tauri Stars (TTS) separated by cTTS and wTTS. Few cTTS are in our sample, but wTTS can be found in nearly all period bins.}
	\label{fig:TTS}
\end{figure}

For an additional regrouping we used all members with observed periods and an H$\alpha$ emission from \cite{luhmanMembers}. Within this sample we classified all stars with an H$\alpha$ equivalent width greater than 10\,\AA{} as classical T-Tauri stars (cTTS) and all other as weak-lined T-Tauri stars (wTTS) \citep{herbig}\footnote{Using the more detailed, spectral type-dependent classification of \cite{TTSclass} would not change our results and we decided to use a single critical value to distinguish between the two groups.}. Stars without an H$\alpha$  equivalent width measurement were excluded from the sample. In \autoref{fig:TTS} we show the histogram of all rotation periods separated into the two groups. Due to the small amount of cTTS in our sample we have not applied statistical test to these data. From \autoref{fig:TTS} we can clearly see that wTTS are among slow and fast rotators. The same result was found in a much larger sample from Orion by \cite{Tanveer}, too. The few cTTS are distributed over the whole range of periods and we cannot find any difference between the two groups.

\subsection{UX Orionis type stars in IC~348}

\begin{figure}
	\includegraphics[width=\columnwidth]{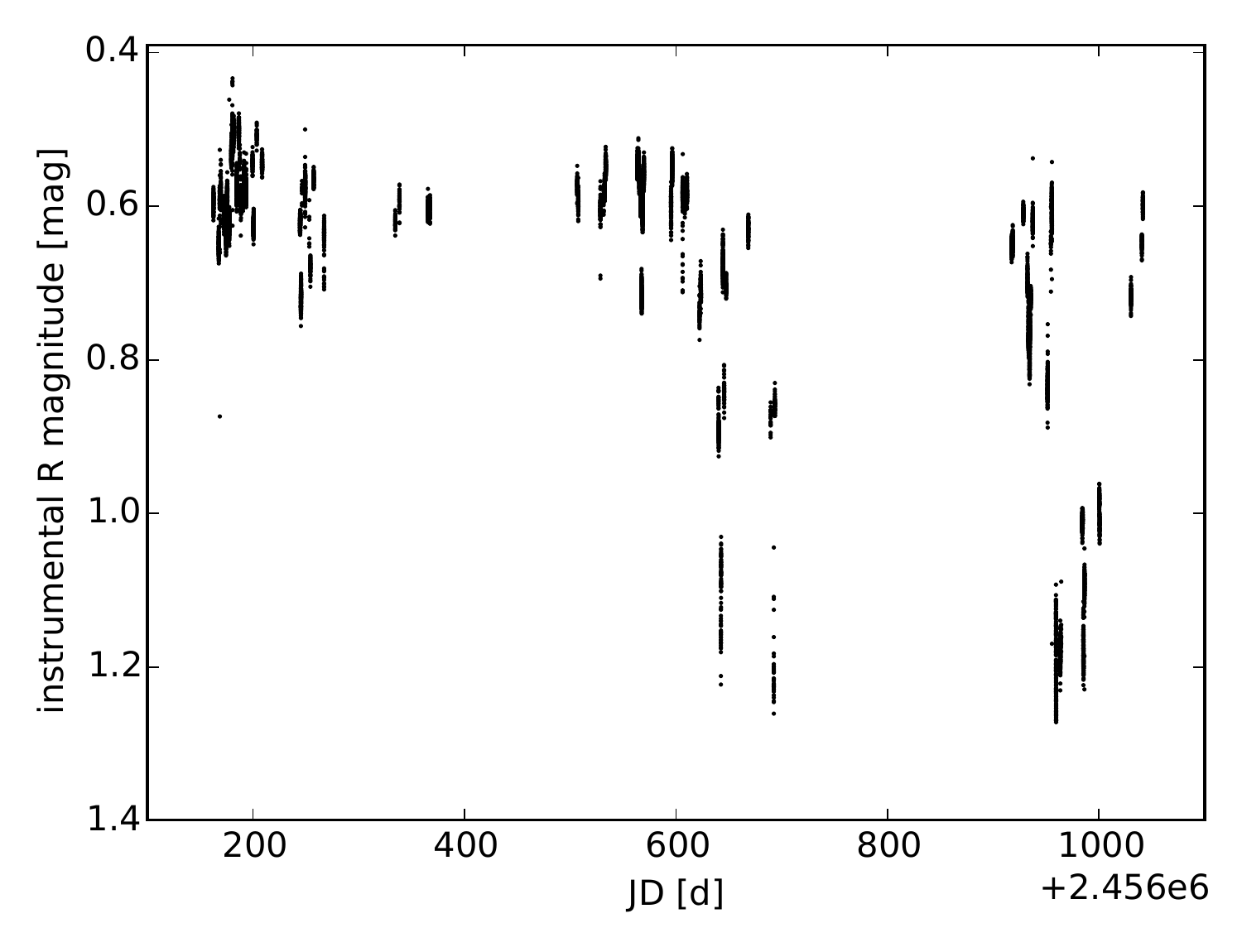}
	\caption{The UX Orionis-type star V909 Per shows sharp, sudden, non-periodic decreases of the flux. This figure includes all observations of this star.}
	\label{fig:78full}
\end{figure}

V909 Per\footnote{FKM 78, LRL 5, HMW 20}, V712 Per\footnote{FKM 140, LRL 37, HMW 23, CB 102}, and V719 Per\footnote{FKM 365, LRL 75, HMW 56, CB 123} have all been classified as UX Orionis-type stars (UXOr) by \cite{uxor}. Those stars are surrounded by clumpy material in their proto-planetary disc. These clumps occults the star which leads to sudden decrease of the flux with few periodicity. From the visual inspection of the light curves we can confirm their results. For V909 Per we show the full light curve in \autoref{fig:78full}. In our light curves we were able to identify the same deep sudden drops as \cite{uxor}. GM Cephei is a similar star which has previously been monitored by YETI \citep{chen}.

While obtaining $BVRI$ photometry V719 Per was covered by its disc and is therefore the faintest star in our sample $R=(18.73\pm0.33)\,\mathrm{mag}$. Additionally, it has a colour of $(V-R)=(2.8\pm0.6)\,\mathrm{mag}$ and is one of the reddest stars examined. This observation seems to disagree with the properties of UXOr variables which are expected to become bluer when eclipsed. \cite{uxor} showed that V719 Per appears bluer near the photometric minimum but redder all other times during the eclipse. Our observations were obtained in the first phase of the occultation and therefore the star appeared redder.

\subsection{Non-member variable stars}

Within this data set we found four detached eclipsing binaries which are background stars to IC~348. They have the identifiers 213 (2MASS J03455377+3226418), 777 (CSS J034539.4+314252), 974 (CB 6), and 975 (CB 17). Only 777 is a previously known binary. The stars 975 and 974 were studied by \cite{cieza}, though incorrect periods were stated and the binarity was not mentioned. Therefore, we can conclude that the binarity of this system was unknown before. Of those four stars all but 213 might be W~UMa type binaries. The minima of 213 have rather different depths, therefore it might be a different type of detached binary.

\begin{figure}
	\includegraphics[width=\columnwidth]{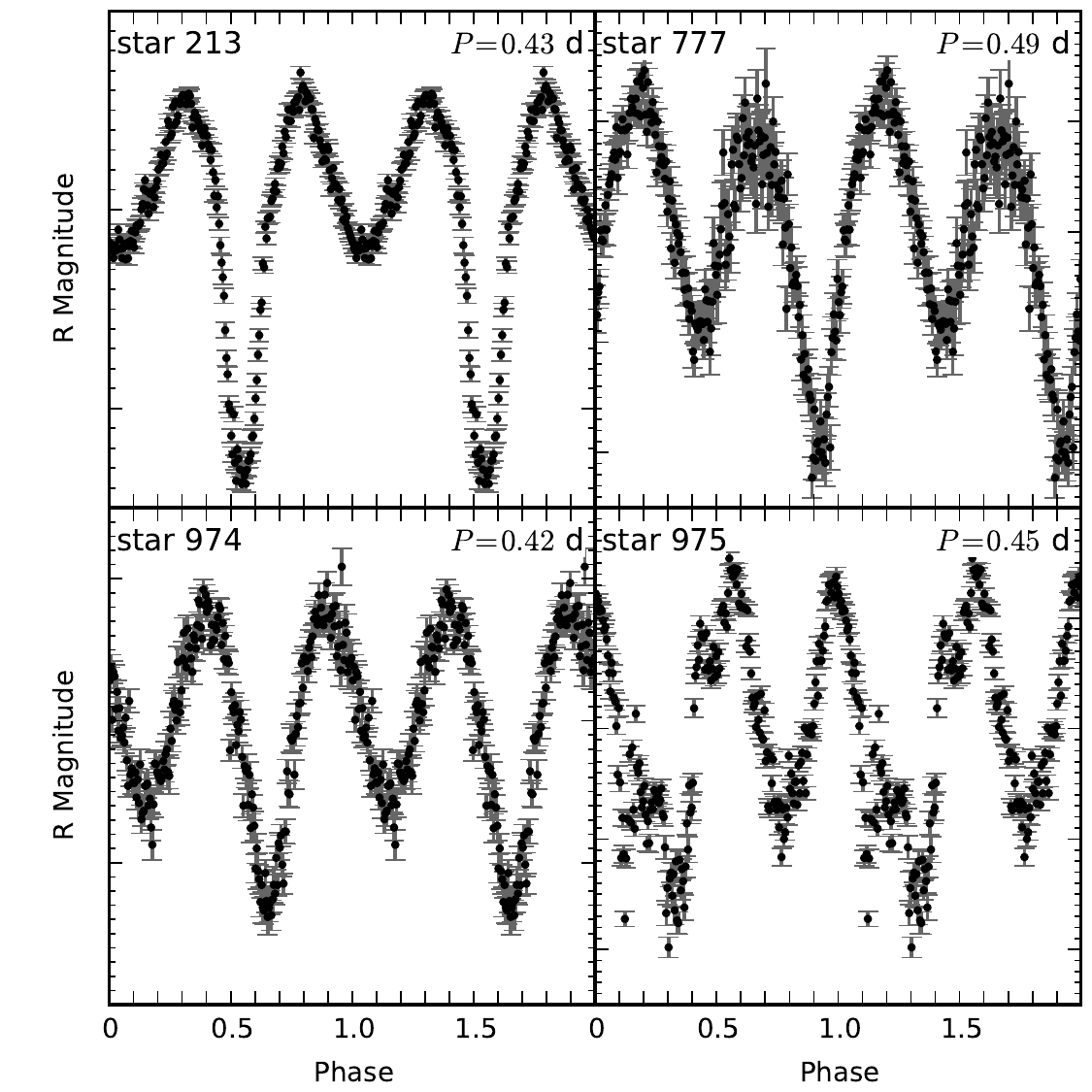}
	\caption{Phase-folded light curves of the four detached binaries in the field of view. Those stars are not members of IC~348. Further information is given in \autoref{tab:binaries}. The spacing is 0.01\,mag for minor ticks and 0.1\,mag for major ticks.}
	\label{fig:binaries}
\end{figure}

With our data we found for 777 a period of $(0.4882\pm 0.0015)\,\mathrm{d}$ which is consistent with the value of 0.48825\,d given by \cite{drake}. The periods of all four binaries are listed in \autoref{tab:binaries} and all four phase-folded light curves are presented \autoref{fig:binaries}.

\begin{table}
	\caption{Periods, brightness, and colours of the four non-member eclipsing binaries found in the data. The magnitudes of the stars 974 and 975 are from \protect\cite{scholz}.}
	\label{tab:binaries}
	\centering
	\begin{tabular}{rccc}
		\hline
		ID & Period & $R$ & ($V-R$)\\
			& [d] & [mag] & [mag]\\
		\hline
		213 & $0.4340\pm0.0015$ & $15.02\pm0.02$ & $0.95\pm0.21$ \\
		777 & $0.4882\pm0.0015$ & $17.26\pm0.10$ & $0.93\pm0.30$ \\
		974 & $0.422\pm0.002$ & $15.41\pm0.02$ & $0.98\pm0.21$ \\
		975 & $0.446\pm0.002$ & $15.72\pm0.02$ & $0.92\pm0.21$\\
		\hline
	\end{tabular}
\end{table}

\begin{figure}
 	\includegraphics[width=\columnwidth]{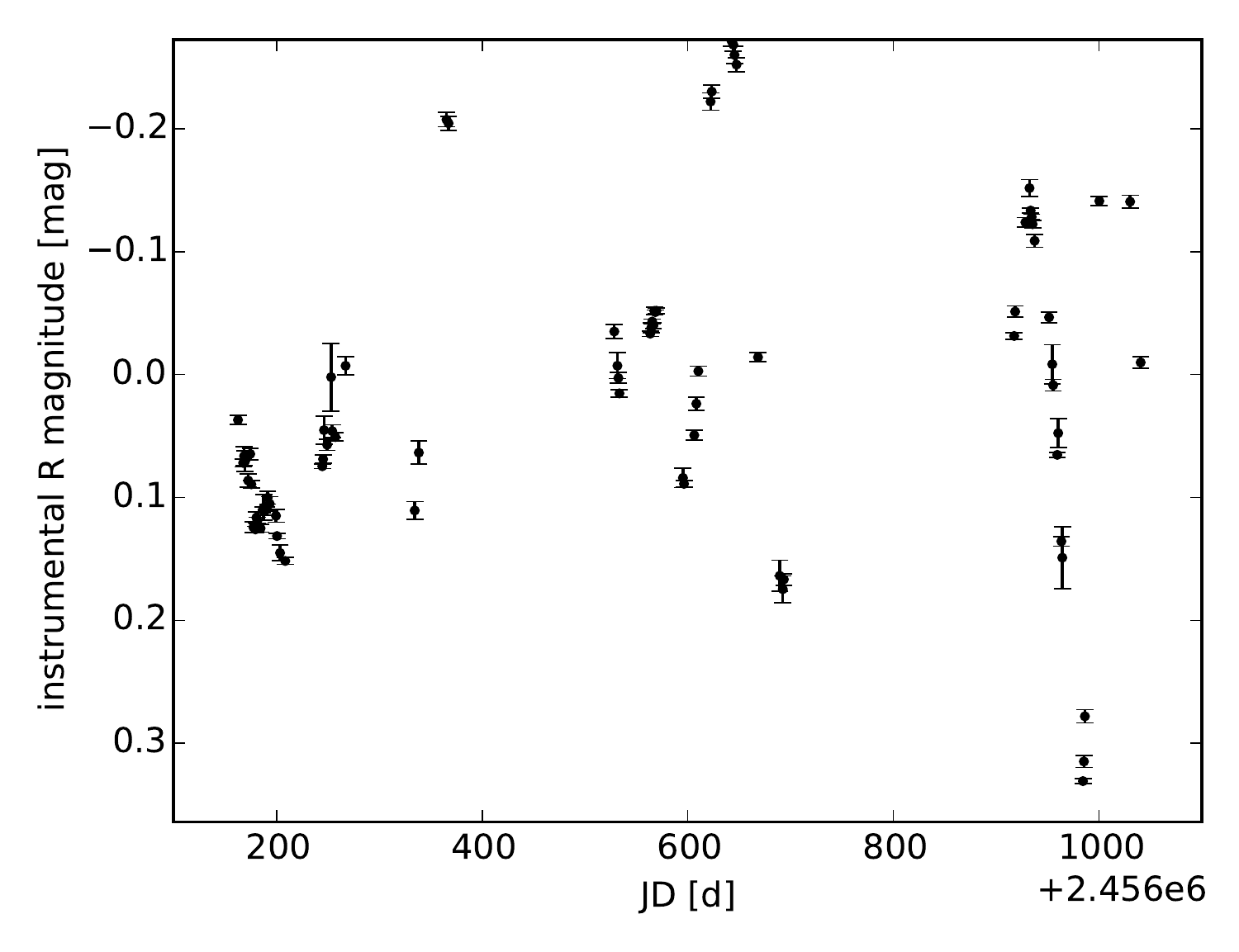}
 	\caption{The full light curve of star 55 (NSVS J0343111+321746) shows quasi-periodic variability. The data have been binned to one data point per night.}
 	\label{fig:55full}
\end{figure}

Among the field stars the star 55 (NSVS J0343111+321746) is a quasi-periodic variable. \cite{wozniak} found a period of 300\,d for this star and classified it as a \textit{SR+L} AGB star. The period cannot be confirmed because we limited our search range to 293\,d. Nevertheless, this star shows clearly a quasi-periodic variability as visible from the light curve in \autoref{fig:55full}. The classification as a \textit{SR+L} AGB star can be confirmed.

\subsection{Stability of periods}

\cite{CohenHerbst} reported that they found not a single variable star in IC~348 with a coherent phase for more than one observational season. In contrast \cite{rebull} found $\sim30$ per cent of the stars in the similar aged flanking fields of ONC to be stable up to one year. In this study we can report coherent phases of the stars in IC~348. Overall 48 per cent of the members have a coherent phase over the range of 2.4\,yr.

This behaviour can be seen in \autoref{fig:lcs} where we show the phase-folded light curves of the photometric periodic IC~348 members. All stars which have a season number added to their name have not been observed in a coherent phase over all three seasons.

\subsection{Colour-Magnitude-Diagram}

\begin{figure}
	\includegraphics[width=\columnwidth]{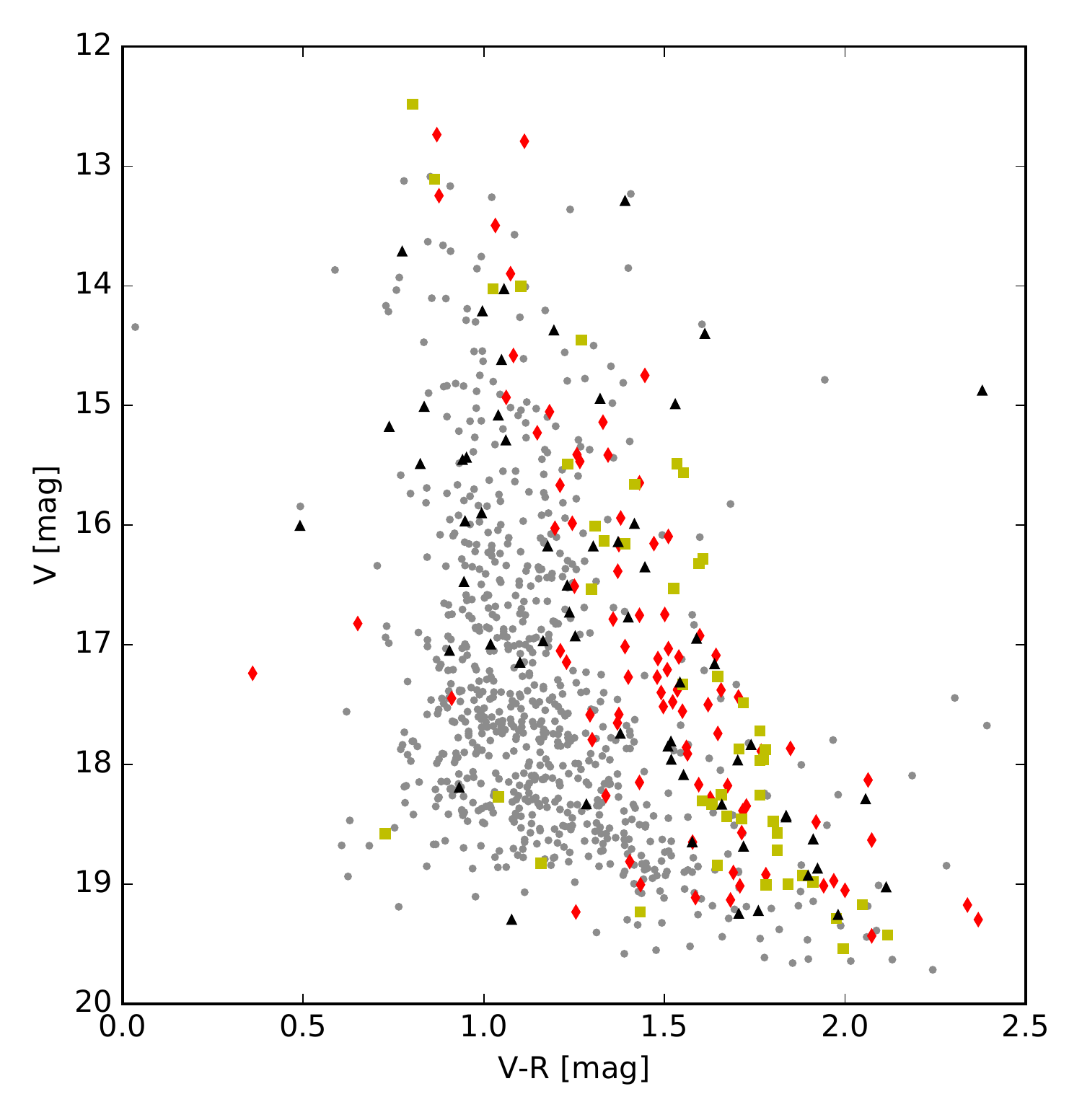}
	\caption{Colour-magnitude-diagram of our field of view. The stars are marked with the same symbols as in \autoref{fig:bvr} (except that the periodic non-members are marked in black instead of white) and the field stars are added with light grey circles. The cluster sequence is clearly visible. The overlap of the cluster sequence with other periodically variable stars is due to the extinction towards IC~348 (see text).}
	\label{fig:cmd}
\end{figure}

In \autoref{fig:cmd} we present the colour-magnitude-diagram (CMD) of the whole FoV as measured from our $V$ and $R$ images. The diagram has not been dereddened and presents the observed colours and magnitudes. In this figure we marked the periodic stars as well as the members of IC~348. The members form a well defined cluster sequence. Some outlier are bluer than expected. Those stars are located close to the bright star Atik (o~Per), which may explain the anomalous colour measurements.

In addition to the members on the cluster sequence a lot of photometric periodic stars, which are not members of the cluster, have similar positions in the CMD. Nevertheless most of those stars have sky positions which do not correspond to the position of IC~348 (\autoref{fig:bvr}). Those stars are reddened by the medium around IC~348 and are background stars. Only two stars on the cluster sequence have a sky position corresponding with IC~348. One of them, LRL~77 (80), has been classified by \cite{luhmanMembers} as a foreground star from proper motion. Our period of $P=15.3\,d$ confirms this classification. This period is too long for a typical member of the cluster. The other one star 607 ([PSZ2003] J034450.0+320345, \citealt{preibisch03}) is positioned on the souther outskirts of the cluster and we find a period of $P=200\,d$. In conclusion, none of the non-members near the sky position of the cluster and on the cluster sequence in the CMD is an unidentified member.

\section[]{Discussion}
\label{sec:dis}

From the long-term photometry of the young open cluster IC~348 we found various periodic variable stars.
For all periods given in \autoref{tab:allperiods} the systematic error is 1 per cent of the given period. This systematic error arises through the selection process of the best period. In our selection algorithm two periods are consistent if they differ by less than 1 per cent. Therefore, the true period can have an error of up to 1 per cent. For short periods the phase difference of small errors is percentually bigger than for long periods. Therefore, a short period can be determined more accurately. In contrast to this strong confinement from the periodogram rotating stars do not exhibit such a well defined periodicity. Due to the evolution of starspots a measurement error is introduced. This error is hard to estimate because it arises from physical processes in the star, with unknown properties for this particular star. Measuring the period of a differentially rotating star can lead to results differing as much as 10 per cent (s. \autoref{sec:compare}). A 1 per cent error is therefore a reasonable estimate even for fast rotating stars.

The periodicity of binary stars can be determined with greater accuracy because the orbital period is not subject to phase changes. If it is possible from the periodograms we give smaller errors for those stars in \autoref{tab:allperiods}.

\subsection{Comparison with previous work}
\label{sec:compare}
In the previous work of \cite{cieza} all known periods have been merged into a single data set. From this base we can compare our results with the known periods.  The comparison with \cite{cieza} shows that our method is reliable. Only six periods found in both works differ strongly. When phase-folding our light curves with the period given by \cite{cieza} we cannot see any coherent behaviour in the light curve. The periods given by \cite{cieza} might suffer from beating with the sampling rate. A very special case is the star 968 (CB 62, LRL 72) for which the authors gave a period of 1\,d which is very likely an alias period, since the observations were carried out in a few nights at one observatory. Our data with a better phase coverage show a periodicity of $(44.6\pm0.5)$\,d.

\begin{figure}
	\includegraphics[width=\columnwidth]{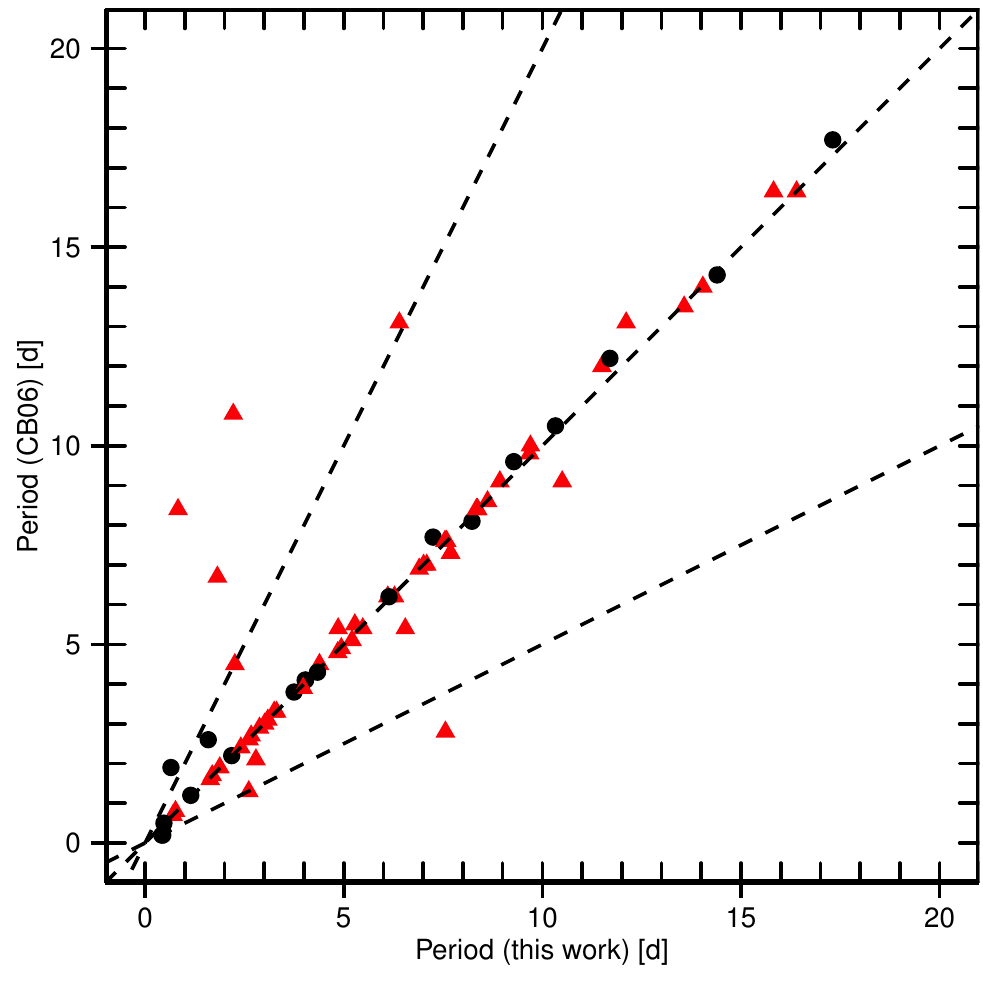}
	\caption{Comparison of our periods to the data of \protect\cite{cieza} (CB06). The (red) triangles represent members of IC~348 and (black) dots are field stars. The dotted lines correspond to the relations of 1:2, 1:1 and 2:1 (from top to bottom).}
	\label{fig:CompCB06}
\end{figure}

For all other stars found in both studies the periods match well with the line of equality, or in some cases have a ratio of 1:2 or 2:1 (\autoref{fig:CompCB06}). Some periods differ by 0.3\,d which can be explained by differentially rotating stars and spot evolution (see \citealt{rebull} and \citealt{cieza}).

For 24 stars for which \cite{cieza} gave a period we were not able to find any periodic variability. Given the activity of the young stars this is not a surprising result. Several reasons can lead to a non-detection. On the one hand the star might have shown no periodic behaviour over the course of three years. Some observed stars were periodic only in one out of three seasons and a lot of stars are not periodic at all, which does not mean that no starspots were present. The stars are still very variable, but no periodicity can be detected because of the number or the evolution of the starspots. For the stars which showed no periodic variability \cite{jackson} found in NGC~2516 no qualitative difference to stars which did. Therefore it is not unusual to miss rotational periods published before. For the same reason it was possible to find 25 new rotation periods although IC~348 is well researched. On the other hand the variability due to spots can easily be covered by other non-periodic behaviours. If the noise in terms of erratic behaviour of the light curve is too strong, even the best algorithms cannot find the underlying period in the light curve.

\cite{cieza} stated the number of periodic stars in IC~348 to be 143. This included 37 stars not listed as members in \cite{luhmanMembers}. With the membership criterion of \cite{cieza} (everything in the FoV of their study, including foreground and background objects) the number of stars in IC~348 which have shown some periodicity is around 200.
  
\subsection{Further periodic variables in the field of view}

The case of LRL~47 shows the importance of the long-term monitoring of young open clusters for finding companions. Although, the periodicity of the observed transit is 5.12\,d it was observed only four times over three observing seasons. Other surveys have missed this event before and we were only able to calculate a period because we used observations distributed through YETI at observatories worldwide.

By analysing the whole FoV and not just the members of the cluster we were able to find additional periodic stars, including three unknown contact binaries. This shows there is additional scientific value in analysing all available data, and not just primary targets (in this case the stars associated with the cluster).

\section[]{Conclusions}

With long-term photometry from different telescopes world-wide we were able to identify 148 periodic variabilities in our FoV centred at IC~348. Because of the long time-scale and the wide FoV 74 of the periods are newly discovered. This includes 33 new periodic variables in IC~348. The total number of periodic stars among the members of IC~348 (counting only the stars of \cite{luhmannew}) rises with this study to 139. Discovering new periods in this well studied cluster is possible due to the high variability of the starspots on these young stars. With a time base large enough it should be possible to determine rotational periods for most stars in young clusters. Photometric long-term studies are therefore viable sources of statistically significant data on the rotation of stars.

With the additional periods for the stars in IC~348 we were able to show that a bimodal Gaussian distribution fits the data more likely than a unimodal. Nevertheless, a dip test was inconclusive and we have no statistical evidence that the period distribution of the stars in IC~348 is bimodal. In addition we showed that for the stars of IC~348 the rotation period does not depend on the type of the T-Tauri stars (TTS). Weak-lined and classical TTS can both be found among fast and slow rotators. The same applies to stars with and without an observed disc indicator. Stars without a disc can also be found among the slow rotators.

For LRL 47 we found a close low-mass stellar companion. The companion has been observed in a grazing transit hence our analysis from single-band photometric data is not reliable. For further characterization of this system -- including mass determination -- radial velocity data are desirable.

Outside the cluster we discovered three previously unknown background detached binaries and were able to confirm the period of a fourth.

\section*{Acknowledgments}
We would like to thank the anonymous referee for the helpful comments which improved and clarified the paper.
We are grateful to D. Cohen and E. Jensen for obtaining valuable observations at the Peter van de Kamp observatory, Swarthmore College and Y.~Oasa at Saitama University Observatory.
We thank A. Dathe, A. Ide, D. Keeley, W. Pfau, S. Schönfeld, and L. Trepl for observing at the University Observatory Jena.
Based on observations obtained with telescopes of the University Observatory Jena, which is operated by the Astrophysical Institute and University-Observatory of the Friedrich-Schiller-University, and on observations obtained at the Llano del Hato National Astronomical Observatory of Venezuela, operated by the Centro de Investigaciones de Astronom{\'\i}a (CIDA) for the Ministerio del Poder Popular para la Ciencia y Tecnolog{\'\i}a.
We would like to thank the Thuringian State (Th\"uringer Ministerium f\"ur Bildung, Wissenschaft und Kultur) in project number B 515-07010 for financial support. 
MK thanks RN and Prof. R. Redmer in DFG projects NE 515 / 34-1 \& 2
and RE 882 / 12-2, respectively, both in SPP 1385.
CA thanks RN in DFG projects NE 515 / 35-1 \& 2 in SPP 1385.
TB and RWW are grateful to the Science and Technology Facilities Committee (STFC) for financial support (grant reference ST/J001236/1).
BD thanks RN in DFG project C2 in SFB-TR 7.
RE thanks RN in DFG projects NE 515 / 34-1 \& 2 in SPP 1385.
CG thanks MM in DFG project MU 2695 / 18-1.
CM thanks K. Schreyer in DFG project SCHR 665 / 7-1.
AP thanks RN in DFG project C2 in SFB-TR 7.
SR thanks RN in DFG projects NE 515 / 33-1 \& 2 in SPP 1385.
SR is currently a Research Fellow at ESA/ESTEC.
JGS thanks RN in DFG project B9 in SFB-TR 7.
TOBS thanks Prof. J. Schmitt in DFG RTG 1351 \textit{Extrasolar Planets and their Host Stars}.
This work was supported by grant VEGA 2/0143/14 of Slovak Academy of Sciences.
IRAF is distributed by the National Optical Astronomy Observatories, which are operated by the Association of Universities for Research in Astronomy, Inc., under cooperative agreement with the National Science Foundation.
This research has made use of NASA's Astrophysics Data System.
This research has made use of the VizieR catalogue access tool and SIMBAD database, both operated at CDS, Strasbourg, France. The original description of the VizieR service was published in A\&AS 143, 23. "The SIMBAD astronomical database" in A\&AS, 143, 9 Wenger et al. 
This research has made use of TOPCAT \citep{topcat} and astrometry.net \citep{astrometry}.
This publication makes use of data products from the Two Micron All Sky Survey, which is a joint project of the University of Massachusetts and the Infrared Processing and Analysis Center/California Institute of Technology, funded by the National Aeronautics and Space Administration and the National Science Foundation.

%%%%%%%%%%%%%%%%%%%%%%%%%%%%%%%%%%%%%%%%%%%%%%%%%%



\bibliographystyle{mnras}
\bibliography{master_ref}

\begin{thebibliography}{}
\makeatletter
\relax
\def\mn@urlcharsother{\let\do\@makeother \do\$\do\&\do\#\do\^\do\_\do\%\do\~}
\def\mn@doi{\begingroup\mn@urlcharsother \@ifnextchar [ {\mn@doi@}
  {\mn@doi@[]}}
\def\mn@doi@[#1]#2{\def\@tempa{#1}\ifx\@tempa\@empty \href
  {http://dx.doi.org/#2} {doi:#2}\else \href {http://dx.doi.org/#2} {#1}\fi
  \endgroup}
\def\mn@eprint#1#2{\mn@eprint@#1:#2::\@nil}
\def\mn@eprint@arXiv#1{\href {http://arxiv.org/abs/#1} {{\tt arXiv:#1}}}
\def\mn@eprint@dblp#1{\href {http://dblp.uni-trier.de/rec/bibtex/#1.xml}
  {dblp:#1}}
\def\mn@eprint@#1:#2:#3:#4\@nil{\def\@tempa {#1}\def\@tempb {#2}\def\@tempc
  {#3}\ifx \@tempc \@empty \let \@tempc \@tempb \let \@tempb \@tempa \fi \ifx
  \@tempb \@empty \def\@tempb {arXiv}\fi \@ifundefined
  {mn@eprint@\@tempb}{\@tempb:\@tempc}{\expandafter \expandafter \csname
  mn@eprint@\@tempb\endcsname \expandafter{\@tempc}}}

\bibitem[\protect\citeauthoryear{{Attridge} \& {Herbst}}{{Attridge} \&
  {Herbst}}{1992}]{ONCherbst}
{Attridge} J.~M.,  {Herbst} W.,  1992, \mn@doi [\apjl] {10.1086/186577}, \href
  {http://adsabs.harvard.edu/abs/1992ApJ...398L..61A} {398, L61}

\bibitem[\protect\citeauthoryear{{Bakos}, {Noyes}, {Kov{\'a}cs}, {Stanek},
  {Sasselov}  \& {Domsa}}{{Bakos} et~al.}{2004}]{hatp}
{Bakos} G.,  {Noyes} R.~W.,  {Kov{\'a}cs} G.,  {Stanek} K.~Z.,  {Sasselov}
  D.~D.,   {Domsa} I.,  2004, \mn@doi [\pasp] {10.1086/382735}, \href
  {http://adsabs.harvard.edu/abs/2004PASP..116..266B} {116, 266}

\bibitem[\protect\citeauthoryear{{Baraffe}, {Homeier}, {Allard}  \&
  {Chabrier}}{{Baraffe} et~al.}{2015}]{BaraffeNew}
{Baraffe} I.,  {Homeier} D.,  {Allard} F.,   {Chabrier} G.,  2015, \mn@doi
  [\aap] {10.1051/0004-6361/201425481}, \href
  {http://adsabs.harvard.edu/abs/2015A%26A...577A..42B} {577, A42}

\bibitem[\protect\citeauthoryear{{Barrado y Navascu{\'e}s} \&
  {Mart{\'{\i}}n}}{{Barrado y Navascu{\'e}s} \&
  {Mart{\'{\i}}n}}{2003}]{TTSclass}
{Barrado y Navascu{\'e}s} D.,  {Mart{\'{\i}}n} E.~L.,  2003, \mn@doi [\aj]
  {10.1086/379673}, \href {http://adsabs.harvard.edu/abs/2003AJ....126.2997B}
  {126, 2997}

\bibitem[\protect\citeauthoryear{{Barsunova}, {Grinin}  \&
  {Sergeev}}{{Barsunova} et~al.}{2013}]{V695}
{Barsunova} O.~Y.,  {Grinin} V.~P.,   {Sergeev} S.~G.,  2013, \mn@doi
  [Astrophysics] {10.1007/s10511-013-9294-5}, \href
  {http://adsabs.harvard.edu/abs/2013Ap.....56..395B} {56, 395}

\bibitem[\protect\citeauthoryear{{Barsunova}, {Grinin}, {Sergeev}, {Semenov}
  \& {Shugarov}}{{Barsunova} et~al.}{2015}]{uxor}
{Barsunova} O.~Y.,  {Grinin} V.~P.,  {Sergeev} S.~G.,  {Semenov} A.~O.,
  {Shugarov} S.~Y.,  2015, \mn@doi [Astrophysics] {10.1007/s10511-015-9375-8},
  \href {http://adsabs.harvard.edu/abs/2015Ap.....58..193B} {58, 193}

\bibitem[\protect\citeauthoryear{{Bell}, {Naylor}, {Mayne}, {Jeffries}  \&
  {Littlefair}}{{Bell} et~al.}{2013}]{Bell}
{Bell} C.~P.~M.,  {Naylor} T.,  {Mayne} N.~J.,  {Jeffries} R.~D.,
  {Littlefair} S.~P.,  2013, \mn@doi [\mnras] {10.1093/mnras/stt1075}, \href
  {http://adsabs.harvard.edu/abs/2013MNRAS.434..806B} {434, 806}

\bibitem[\protect\citeauthoryear{Benaglia, Chauveau, Hunter  \& Young}{Benaglia
  et~al.}{2009}]{mixtools}
Benaglia T.,  Chauveau D.,  Hunter D.~R.,   Young D.,  2009, Journal of
  Statistical Software, 32, 1

\bibitem[\protect\citeauthoryear{{Bord{\'e}}, {Rouan}  \&
  {L{\'e}ger}}{{Bord{\'e}} et~al.}{2003}]{corot}
{Bord{\'e}} P.,  {Rouan} D.,   {L{\'e}ger} A.,  2003, \mn@doi [\aap]
  {10.1051/0004-6361:20030675}, \href
  {http://adsabs.harvard.edu/abs/2003A%26A...405.1137B} {405, 1137}

\bibitem[\protect\citeauthoryear{{Broeg}, {Fern{\'a}ndez}  \&
  {Neuhäuser}}{{Broeg} et~al.}{2005}]{broeg}
{Broeg} C.,  {Fern{\'a}ndez} M.,   {Neuhäuser} R.,  2005, \mn@doi
  [Astronomische Nachrichten] {10.1002/asna.200410350}, \href
  {http://adsabs.harvard.edu/abs/2005AN....326..134B} {326, 134}

\bibitem[\protect\citeauthoryear{{Chen} et~al.,}{{Chen} et~al.}{2012}]{chen}
{Chen} W.~P.,  et~al., 2012, \mn@doi [\apj] {10.1088/0004-637X/751/2/118},
  \href {http://adsabs.harvard.edu/abs/2012ApJ...751..118C} {751, 118}

\bibitem[\protect\citeauthoryear{{Cieza} \& {Baliber}}{{Cieza} \&
  {Baliber}}{2006}]{cieza}
{Cieza} L.,  {Baliber} N.,  2006, \mn@doi [\apj] {10.1086/506342}, \href
  {http://cdsads.u-strasbg.fr/abs/2006ApJ...649..862C} {649, 862}

\bibitem[\protect\citeauthoryear{{Cieza} et~al.,}{{Cieza}
  et~al.}{2007}]{cieza07}
{Cieza} L.,  et~al., 2007, \mn@doi [\apj] {10.1086/520698}, \href
  {http://adsabs.harvard.edu/abs/2007ApJ...667..308C} {667, 308}

\bibitem[\protect\citeauthoryear{{Cohen}, {Herbst}  \& {Williams}}{{Cohen}
  et~al.}{2003}]{CohenUnusual}
{Cohen} R.~E.,  {Herbst} W.,   {Williams} E.~C.,  2003, \mn@doi [\apjl]
  {10.1086/379275}, \href {http://adsabs.harvard.edu/abs/2003ApJ...596L.243C}
  {596, L243}

\bibitem[\protect\citeauthoryear{{Cohen}, {Herbst}  \& {Williams}}{{Cohen}
  et~al.}{2004}]{CohenHerbst}
{Cohen} R.~E.,  {Herbst} W.,   {Williams} E.~C.,  2004, \mn@doi [\aj]
  {10.1086/381925}, \href {http://adsabs.harvard.edu/abs/2004AJ....127.1602C}
  {127, 1602}

\bibitem[\protect\citeauthoryear{{Currie} \& {Kenyon}}{{Currie} \&
  {Kenyon}}{2009}]{currie}
{Currie} T.,  {Kenyon} S.~J.,  2009, \mn@doi [\aj]
  {10.1088/0004-6256/138/3/703}, \href
  {http://cdsads.u-strasbg.fr/abs/2009AJ....138..703C} {138, 703}

\bibitem[\protect\citeauthoryear{{Dahm}}{{Dahm}}{2008}]{dahm}
{Dahm} S.~E.,  2008, \mn@doi [\aj] {10.1088/0004-6256/136/2/521}, \href
  {http://adsabs.harvard.edu/abs/2008AJ....136..521D} {136, 521}

\bibitem[\protect\citeauthoryear{{Drake} et~al.,}{{Drake} et~al.}{2014}]{drake}
{Drake} A.~J.,  et~al., 2014, \mn@doi [\apjs] {10.1088/0067-0049/213/1/9},
  \href {http://adsabs.harvard.edu/abs/2014ApJS..213....9D} {213, 9}

\bibitem[\protect\citeauthoryear{{Duch{\^e}ne}, {Bouvier}  \&
  {Simon}}{{Duch{\^e}ne} et~al.}{1999}]{duchene}
{Duch{\^e}ne} G.,  {Bouvier} J.,   {Simon} T.,  1999, \aap, \href
  {http://cdsads.u-strasbg.fr/abs/1999A%26A...343..831D} {343, 831}

\bibitem[\protect\citeauthoryear{{Dworetsky}}{{Dworetsky}}{1983}]{dworetsky}
{Dworetsky} M.~M.,  1983, \mnras, \href
  {http://adsabs.harvard.edu/abs/1983MNRAS.203..917D} {203, 917}

\bibitem[\protect\citeauthoryear{{Errmann} et~al.,}{{Errmann}
  et~al.}{2014}]{errmann}
{Errmann} R.,  et~al., 2014, \mn@doi [Astronomische Nachrichten]
  {10.1002/asna.201412047}, \href
  {http://adsabs.harvard.edu/abs/2014AN....335..345E} {335, 345}

\bibitem[\protect\citeauthoryear{{Etzel}}{{Etzel}}{1981}]{Etzel}
{Etzel} P.~B.,  1981, in {Carling} E.~B.,  {Kopal} Z.,  eds, Photometric and
  Spectroscopic Binary Systems. p.~111

\bibitem[\protect\citeauthoryear{{Flaherty} et~al.,}{{Flaherty}
  et~al.}{2014}]{FlahertyXray}
{Flaherty} K.~M.,  et~al., 2014, \mn@doi [\apj] {10.1088/0004-637X/793/1/2},
  \href {http://cdsads.u-strasbg.fr/abs/2014ApJ...793....2F} {793, 2}

\bibitem[\protect\citeauthoryear{{Garai} et~al.,}{{Garai}
  et~al.}{2016}]{garaiYETI}
{Garai} Z.,  et~al., 2016, \mn@doi [Astronomische Nachrichten]
  {10.1002/asna.201512310}, \href
  {http://adsabs.harvard.edu/abs/2016AN....337..261G} {337, 261}

\bibitem[\protect\citeauthoryear{{Gregory}}{{Gregory}}{1999}]{gregory99}
{Gregory} P.~C.,  1999, \mn@doi [\apj] {10.1086/307433}, \href
  {http://adsabs.harvard.edu/abs/1999ApJ...520..361G} {520, 361}

\bibitem[\protect\citeauthoryear{{Gregory} \& {Loredo}}{{Gregory} \&
  {Loredo}}{1992}]{gregory}
{Gregory} P.~C.,  {Loredo} T.~J.,  1992, \mn@doi [\apj] {10.1086/171844}, \href
  {http://adsabs.harvard.edu/abs/1992ApJ...398..146G} {398, 146}

\bibitem[\protect\citeauthoryear{{Grinin}, {Stempels}, {Gahm}, {Sergeev},
  {Arkharov}, {Barsunova}  \& {Tambovtseva}}{{Grinin} et~al.}{2008}]{Grinin}
{Grinin} V.~P.,  {Stempels} H.~C.,  {Gahm} G.~F.,  {Sergeev} S.,  {Arkharov}
  A.,  {Barsunova} O.,   {Tambovtseva} L.,  2008, \mn@doi [\aap]
  {10.1051/0004-6361:200810349}, \href
  {http://adsabs.harvard.edu/abs/2008A%26A...489.1233G} {489, 1233}

\bibitem[\protect\citeauthoryear{{Hardy}, {Butterley}, {Dhillon}, {Littlefair}
  \& {Wilson}}{{Hardy} et~al.}{2015}]{pt5m}
{Hardy} L.~K.,  {Butterley} T.,  {Dhillon} V.~S.,  {Littlefair} S.~P.,
  {Wilson} R.~W.,  2015, \mn@doi [\mnras] {10.1093/mnras/stv2279}, \href
  {http://adsabs.harvard.edu/abs/2015MNRAS.454.4316H} {454, 4316}

\bibitem[\protect\citeauthoryear{Hartigan}{Hartigan}{1985}]{hartigan}
Hartigan P.~M.,  1985, Journal of the Royal Statistical Society. Series C
  (Applied Statistics), 34, 320

\bibitem[\protect\citeauthoryear{Hartigan \& Hartigan}{Hartigan \&
  Hartigan}{1985}]{hartigan2}
Hartigan J.~A.,  Hartigan P.~M.,  1985, The Annals of Statistics, 13, 70

\bibitem[\protect\citeauthoryear{{Herbig}}{{Herbig}}{1954}]{HerbigTTS}
{Herbig} G.~H.,  1954, \mn@doi [\pasp] {10.1086/126641}, \href
  {http://adsabs.harvard.edu/abs/1954PASP...66...19H} {66, 19}

\bibitem[\protect\citeauthoryear{{Herbig}}{{Herbig}}{1998}]{herbig}
{Herbig} G.~H.,  1998, \mn@doi [\apj] {10.1086/305500}, \href
  {http://adsabs.harvard.edu/abs/1998ApJ...497..736H} {497, 736}

\bibitem[\protect\citeauthoryear{{Herbst}, {Rhode}, {Hillenbrand}  \&
  {Curran}}{{Herbst} et~al.}{2000a}]{herbstONC2000}
{Herbst} W.,  {Rhode} K.~L.,  {Hillenbrand} L.~A.,   {Curran} G.,  2000a,
  \mn@doi [\aj] {10.1086/301175}, \href
  {http://adsabs.harvard.edu/abs/2000AJ....119..261H} {119, 261}

\bibitem[\protect\citeauthoryear{{Herbst}, {Maley}  \& {Williams}}{{Herbst}
  et~al.}{2000b}]{herbst}
{Herbst} W.,  {Maley} J.~A.,   {Williams} E.~C.,  2000b, \mn@doi [\aj]
  {10.1086/301430}, \href {http://adsabs.harvard.edu/abs/2000AJ....120..349H}
  {120, 349}

\bibitem[\protect\citeauthoryear{{Jackson} \& {Jeffries}}{{Jackson} \&
  {Jeffries}}{2012}]{jackson}
{Jackson} R.~J.,  {Jeffries} R.~D.,  2012, \mn@doi [\mnras]
  {10.1111/j.1365-2966.2012.21119.x}, \href
  {http://adsabs.harvard.edu/abs/2012MNRAS.423.2966J} {423, 2966}

\bibitem[\protect\citeauthoryear{{Jeffries}, {Littlefair}, {Naylor}  \&
  {Mayne}}{{Jeffries} et~al.}{2011}]{jeffriesONC}
{Jeffries} R.~D.,  {Littlefair} S.~P.,  {Naylor} T.,   {Mayne} N.~J.,  2011,
  \mn@doi [\mnras] {10.1111/j.1365-2966.2011.19613.x}, \href
  {http://adsabs.harvard.edu/abs/2011MNRAS.418.1948J} {418, 1948}

\bibitem[\protect\citeauthoryear{{Lada} et~al.,}{{Lada} et~al.}{2006}]{lada}
{Lada} C.~J.,  et~al., 2006, \mn@doi [\aj] {10.1086/499808}, \href
  {http://cdsads.u-strasbg.fr/abs/2006AJ....131.1574L} {131, 1574}

\bibitem[\protect\citeauthoryear{{Lang}, {Hogg}, {Mierle}, {Blanton}  \&
  {Roweis}}{{Lang} et~al.}{2010}]{astrometry}
{Lang} D.,  {Hogg} D.~W.,  {Mierle} K.,  {Blanton} M.,   {Roweis} S.,  2010,
  \mn@doi [\aj] {10.1088/0004-6256/139/5/1782}, \href
  {http://adsabs.harvard.edu/abs/2010AJ....139.1782L} {139, 1782}

\bibitem[\protect\citeauthoryear{{Littlefair}, {Naylor}, {Burningham}  \&
  {Jeffries}}{{Littlefair} et~al.}{2005}]{littlefair}
{Littlefair} S.~P.,  {Naylor} T.,  {Burningham} B.,   {Jeffries} R.~D.,  2005,
  \mn@doi [\mnras] {10.1111/j.1365-2966.2005.08737.x}, \href
  {http://adsabs.harvard.edu/abs/2005MNRAS.358..341L} {358, 341}

\bibitem[\protect\citeauthoryear{{Luhman}, {Rieke}, {Lada}  \& {Lada}}{{Luhman}
  et~al.}{1998}]{luhmanIMF}
{Luhman} K.~L.,  {Rieke} G.~H.,  {Lada} C.~J.,   {Lada} E.~A.,  1998, \mn@doi
  [\apj] {10.1086/306393}, \href
  {http://adsabs.harvard.edu/abs/1998ApJ...508..347L} {508, 347}

\bibitem[\protect\citeauthoryear{{Luhman}, {Stauffer}, {Muench}, {Rieke},
  {Lada}, {Bouvier}  \& {Lada}}{{Luhman} et~al.}{2003}]{luhmanMembers}
{Luhman} K.~L.,  {Stauffer} J.~R.,  {Muench} A.~A.,  {Rieke} G.~H.,  {Lada}
  E.~A.,  {Bouvier} J.,   {Lada} C.~J.,  2003, \mn@doi [\apj] {10.1086/376594},
  \href {http://adsabs.harvard.edu/abs/2003ApJ...593.1093L} {593, 1093}

\bibitem[\protect\citeauthoryear{{Luhman}, {Esplin}  \& {Loutrel}}{{Luhman}
  et~al.}{2016}]{luhmannew}
{Luhman} K.~L.,  {Esplin} T.~L.,   {Loutrel} N.~P.,  2016, preprint (\mn@eprint
  {arXiv} {1605.08907})

\bibitem[\protect\citeauthoryear{Maechler}{Maechler}{2015}]{diptest}
Maechler M.,  2015, diptest: Hartigan's Dip Test Statistic for Unimodality -
  Corrected.
\url {http://CRAN.R-project.org/package=diptest}

\bibitem[\protect\citeauthoryear{{Mugrauer}}{{Mugrauer}}{2016}]{MugrauerCTKII}
{Mugrauer} M.,  2016, \mn@doi [Astronomische Nachrichten]
  {10.1002/asna.201512302}, \href
  {http://adsabs.harvard.edu/abs/2016AN....337..226M} {337, 226}

\bibitem[\protect\citeauthoryear{{Mugrauer} \& {Berthold}}{{Mugrauer} \&
  {Berthold}}{2010}]{MugrauerSTK}
{Mugrauer} M.,  {Berthold} T.,  2010, \mn@doi [Astronomische Nachrichten]
  {10.1002/asna.201011349}, \href
  {http://adsabs.harvard.edu/abs/2010AN....331..449M} {331, 449}

\bibitem[\protect\citeauthoryear{{Neuh{\"a}user}}{{Neuh{\"a}user}}{1997}]{NeuhauserSci}
{Neuh{\"a}user} R.,  1997, \mn@doi [Science] {10.1126/science.276.5317.1363},
  \href {http://adsabs.harvard.edu/abs/1997Sci...276.1363N} {276, 1363}

\bibitem[\protect\citeauthoryear{{Neuh\"auser} et~al.,}{{Neuh\"auser}
  et~al.}{2011}]{NeuhauserYETI}
{Neuh\"auser} R.,  et~al., 2011, \mn@doi [Astronomische Nachrichten]
  {10.1002/asna.201111573}, \href
  {http://adsabs.harvard.edu/abs/2011AN....332..547N} {332, 547}

\bibitem[\protect\citeauthoryear{{Nordhagen}, {Herbst}, {Rhode}  \&
  {Williams}}{{Nordhagen} et~al.}{2006a}]{nordhagen}
{Nordhagen} S.,  {Herbst} W.,  {Rhode} K.~L.,   {Williams} E.~C.,  2006a,
  \mn@doi [\aj] {10.1086/506985}, \href
  {http://adsabs.harvard.edu/abs/2006AJ....132.1555N} {132, 1555}

\bibitem[\protect\citeauthoryear{{Nordhagen}, {Herbst}, {Williams}  \&
  {Semkov}}{{Nordhagen} et~al.}{2006b}]{nordUnusal}
{Nordhagen} S.,  {Herbst} W.,  {Williams} E.~C.,   {Semkov} E.,  2006b, \mn@doi
  [\apjl] {10.1086/507028}, \href
  {http://adsabs.harvard.edu/abs/2006ApJ...646L.151N} {646, L151}

\bibitem[\protect\citeauthoryear{{Nutzman} \& {Charbonneau}}{{Nutzman} \&
  {Charbonneau}}{2008}]{mearth}
{Nutzman} P.,  {Charbonneau} D.,  2008, \mn@doi [\pasp] {10.1086/533420}, \href
  {http://adsabs.harvard.edu/abs/2008PASP..120..317N} {120, 317}

\bibitem[\protect\citeauthoryear{{Popper} \& {Etzel}}{{Popper} \&
  {Etzel}}{1981}]{PopperEtzel}
{Popper} D.~M.,  {Etzel} P.~B.,  1981, \mn@doi [\aj] {10.1086/112862}, \href
  {http://adsabs.harvard.edu/abs/1981AJ.....86..102P} {86, 102}

\bibitem[\protect\citeauthoryear{{Preibisch} \& {Zinnecker}}{{Preibisch} \&
  {Zinnecker}}{2001}]{PZ01}
{Preibisch} T.,  {Zinnecker} H.,  2001, \mn@doi [\aj] {10.1086/321177}, \href
  {http://cdsads.u-strasbg.fr/abs/2001AJ....122..866P} {122, 866}

\bibitem[\protect\citeauthoryear{{Preibisch} \& {Zinnecker}}{{Preibisch} \&
  {Zinnecker}}{2002}]{PZ02}
{Preibisch} T.,  {Zinnecker} H.,  2002, \mn@doi [\aj] {10.1086/338851}, \href
  {http://cdsads.u-strasbg.fr/abs/2002AJ....123.1613P} {123, 1613}

\bibitem[\protect\citeauthoryear{{Preibisch}, {Zinnecker}  \&
  {Herbig}}{{Preibisch} et~al.}{1996}]{preibisch}
{Preibisch} T.,  {Zinnecker} H.,   {Herbig} G.~H.,  1996, \aap, \href
  {http://cdsads.u-strasbg.fr/abs/1996A%26A...310..456P} {310, 456}

\bibitem[\protect\citeauthoryear{{Preibisch}, {Stanke}  \&
  {Zinnecker}}{{Preibisch} et~al.}{2003}]{preibisch03}
{Preibisch} T.,  {Stanke} T.,   {Zinnecker} H.,  2003, \mn@doi [\aap]
  {10.1051/0004-6361:20030973}, \href
  {http://cdsads.u-strasbg.fr/abs/2003A%26A...409..147P} {409, 147}

\bibitem[\protect\citeauthoryear{{R Core Team}}{{R Core Team}}{2015}]{R}
{R Core Team} 2015, R: A Language and Environment for Statistical Computing.
R Foundation for Statistical Computing, Vienna, Austria, \url
  {http://www.R-project.org/}

\bibitem[\protect\citeauthoryear{{Rebull}}{{Rebull}}{2001}]{rebull}
{Rebull} L.~M.,  2001, \mn@doi [\aj] {10.1086/319393}, \href
  {http://adsabs.harvard.edu/abs/2001AJ....121.1676R} {121, 1676}

\bibitem[\protect\citeauthoryear{{Ricker} et~al.,}{{Ricker}
  et~al.}{2014}]{TESS}
{Ricker} G.~R.,  et~al., 2014, in Space Telescopes and Instrumentation 2014:
  Optical, Infrared, and Millimeter Wave. p. 914320 (\mn@eprint {arXiv}
  {1406.0151}), \mn@doi{10.1117/12.2063489}

\bibitem[\protect\citeauthoryear{{Scholz} et~al.,}{{Scholz}
  et~al.}{1999}]{scholz}
{Scholz} R.-D.,  et~al., 1999, \mn@doi [\aaps] {10.1051/aas:1999249}, \href
  {http://adsabs.harvard.edu/abs/1999A%26AS..137..305S} {137, 305}

\bibitem[\protect\citeauthoryear{{Siess}, {Dufour}  \& {Forestini}}{{Siess}
  et~al.}{2000}]{siess}
{Siess} L.,  {Dufour} E.,   {Forestini} M.,  2000, \aap, \href
  {http://adsabs.harvard.edu/abs/2000A%26A...358..593S} {358, 593}

\bibitem[\protect\citeauthoryear{{Southworth}, {Maxted}  \&
  {Smalley}}{{Southworth} et~al.}{2004}]{southworth}
{Southworth} J.,  {Maxted} P.~F.~L.,   {Smalley} B.,  2004, \mn@doi [\mnras]
  {10.1111/j.1365-2966.2004.07871.x}, \href
  {http://adsabs.harvard.edu/abs/2004MNRAS.351.1277S} {351, 1277}

\bibitem[\protect\citeauthoryear{{Stelzer}, {Preibisch}, {Alexander},
  {Mucciarelli}, {Flaccomio}, {Micela}  \& {Sciortino}}{{Stelzer}
  et~al.}{2012}]{Stelzer}
{Stelzer} B.,  {Preibisch} T.,  {Alexander} F.,  {Mucciarelli} P.,  {Flaccomio}
  E.,  {Micela} G.,   {Sciortino} S.,  2012, \mn@doi [\aap]
  {10.1051/0004-6361/201118118}, \href
  {http://cdsads.u-strasbg.fr/abs/2012A%26A...537A.135S} {537, A135}

\bibitem[\protect\citeauthoryear{{Tanveer Karim} et~al.,}{{Tanveer Karim}
  et~al.}{2016}]{Tanveer}
{Tanveer Karim} M.,  et~al., 2016, preprint, \href
  {http://adsabs.harvard.edu/abs/2016arXiv160504333T} {} (\mn@eprint {arXiv}
  {1605.04333})

\bibitem[\protect\citeauthoryear{{Taylor}}{{Taylor}}{2005}]{topcat}
{Taylor} M.~B.,  2005, in {Shopbell} P.,  {Britton} M.,   {Ebert} R.,  eds,
  Astronomical Society of the Pacific Conference Series Vol. 347, Astronomical
  Data Analysis Software and Systems XIV. p.~29

\bibitem[\protect\citeauthoryear{{Trullols} \& {Jordi}}{{Trullols} \&
  {Jordi}}{1997}]{trullols}
{Trullols} E.,  {Jordi} C.,  1997, \aap, \href
  {http://adsabs.harvard.edu/abs/1997A%26A...324..549T} {324, 549}

\bibitem[\protect\citeauthoryear{Venables \& Ripley}{Venables \&
  Ripley}{2002}]{RMASS}
Venables W.~N.,  Ripley B.~D.,  2002, Modern Applied Statistics with S, fourth
  edn.
Springer, New York, \url {http://www.stats.ox.ac.uk/pub/MASS4}

\bibitem[\protect\citeauthoryear{{Wo{\'z}niak}, {Williams}, {Vestrand}  \&
  {Gupta}}{{Wo{\'z}niak} et~al.}{2004}]{wozniak}
{Wo{\'z}niak} P.~R.,  {Williams} S.~J.,  {Vestrand} W.~T.,   {Gupta} V.,  2004,
  \mn@doi [\aj] {10.1086/425526}, \href
  {http://adsabs.harvard.edu/abs/2004AJ....128.2965W} {128, 2965}

\bibitem[\protect\citeauthoryear{{Zechmeister} \& {K{\"u}rster}}{{Zechmeister}
  \& {K{\"u}rster}}{2009}]{zechmeister}
{Zechmeister} M.,  {K{\"u}rster} M.,  2009, \mn@doi [\aap]
  {10.1051/0004-6361:200811296}, \href
  {http://adsabs.harvard.edu/abs/2009A%26A...496..577Z} {496, 577}

\makeatother
\end{thebibliography}

%%%%%%%%%%%%%%%%%%%%%%%%%%%%%%%%%%%%%%%%%%%%%%%%%%

%%%%%%%%%%%%%%%%% APPENDICES %%%%%%%%%%%%%%%%%%%%%

\appendix

\section{Complete list of results}
In the appendix we present a table of our results for all periodically variable stars in the FoV and the phase-folded light curves for all periodically variable stars among the IC~348 members.

\begin{table*}
	\caption{Here we present all periods (members and non-members) found in our data. \textit{ID} gives the number used in this paper, \textit{Identifier} is a name from 2MASS or \protect\cite{preibisch03}. \textit{CB} and \textit{LRL} are the IDs of \protect\cite{cieza} and \protect\cite{luhmanIMF}, respectively. Stars with an LRL ID in parenthesis are non-members of IC 348 according to \protect\cite{luhmanMembers}, other stars with an LRL ID are members from \protect\cite{luhmannew}. $P$ states our period with the error $\Delta P$ while $P_\mathrm{CB}$ is the value given by \protect\cite{cieza}. \textit{Amplitude} gives the peak-to-peak amplitude of the periodic light curve in magnitudes. $R$ and $(V-R)$ are in the Bessel system and have been measured from our data.}
	\label{tab:allperiods}
\begin{tabular}{rrrlrrrrrrrr}
	\hline
	ID & RA & DEC & Identifier & CB & LRL & $P$ & $\Delta P$ & $P_\mathrm{CB}$ & Amplitude & $R$ & $(V-R)$\\
	 & [\degr] & [\degr] & & & & [d] & [d] & [d] & [mag] & [mag] & [mag] \\
	\hline
  25 & 56.1632 & 32.1552 & 2MASS J03443916+3209182 & 109 & 9 & 1.64 & 0.02 & 1.6 & 0.094 & 13.3 & 1.4\\
  27 & 56.1539 & 32.1127 & 2MASS J03443694+3206453 & 95 & 6 & 1.69 & 0.02 & 1.7 & 0.072 & 11.7 & 1.1\\
  30 & 55.9635 & 32.2193 & 2MASS J03435123+3213091 &  & 22 & 0.788 & 0.003 &  & 0.12 & 11.9 & 0.9\\
  31 & 55.8851 & 32.5134 & 2MASS J03433241+3230477 &  &  & 0.78 & 0.01 &  & 0.019 & 11.9 & 1.4\\
  36 & 56.1000 & 32.1834 & 2MASS J03442398+3211000 &  & 38 & 0.71 & 0.01 &  & 0.027 & 12.4 & 0.9\\
  38 & 55.7496 & 32.1541 & 2MASS J03425992+3209144 &  &  & 45.5 & 0.5 &  & 0.21 & 12.5 & 2.4\\
  49 & 55.5409 & 32.3859 & 2MASS J03420982+3223086 &  &  & 39.1 & 0.4 &  & 0.16 & 12.8 & 1.6\\
  52 & 55.5155 & 32.2742 & 2MASS J03420373+3216269 &  &  & 0.1132 & 0.001 &  & 0.012 & 12.9 & 0.8\\
  54 & 55.4924 & 32.2384 & 2MASS J03415816+3214179 &  &  & 4.13 & 0.04 &  & 0.10 & 13.0 & 1.1\\
  60 & 55.8956 & 32.5271 & 2MASS J03433494+3231372 &  &  & 0.84 & 0.01 &  & 0.055 & 13.2 & 1.2\\
  67 & 56.0684 & 32.1654 & 2MASS J03441642+3209552 & 45 & 53 & 3.01 & 0.03 & 3 & 0.15 & 13.2 & 1.0\\
  71 & 55.9813 & 32.1590 & 2MASS J03435550+3209321 &  & 47 & 4.91 & 0.05 &  & 0.88 & 13.5 & 1.1\\
  72 & 55.7439 & 32.4100 & 2MASS J03425852+3224359 &  &  & 32.7 & 0.3 &  & 0.15 & 13.5 & 1.5\\
  80 & 56.1809 & 32.1382 & 2MASS J03444342+3208172 &  & (77) & 15.3 & 0.2 &  & 0.077 & 13.6 & 1.0\\
  84 & 55.4940 & 32.0440 & 2MASS J03415855+3202379 &  &  & 11.51 & 0.12 &  & 0.088 & 13.6 & 1.3\\
  87 & 56.1462 & 32.1270 & 2MASS J03443503+3207370 & 91 & 24 A & 2.26 & 0.02 & 4.5 & 0.12 & 13.8 & 1.3\\
  95 & 56.0209 & 32.1650 & 2MASS J03440499+3209537 &  & 56 & 21.5 & 0.2 &  & 0.063 & 13.9 & 1.2\\
  98 & 56.2563 & 32.1810 & 2MASS J03450151+3210512 & 133 & 79 & 1.88 & 0.02 & 1.9 & 0.071 & 13.9 & 1.1\\
  107 & 56.1452 & 32.1093 & 2MASS J03443487+3206337 & 89 & 48 & 5.48 & 0.05 & 5.4 & 0.15 & 14.2 & 1.3\\
  109 & 55.7419 & 32.3678 & 2MASS J03425806+3222042 &  &  & 28.1 & 0.3 &  & 0.028 & 14.0 & 1.0\\
  110 & 56.0369 & 32.2697 & 2MASS J03440885+3216105 &  & 44 & 7.89 & 0.08 &  & 0.053 & 14.1 & 1.3\\
  112 & 56.3775 & 32.0322 & 2MASS J03453061+3201557 &  & 10363 & 2.24 & 0.02 &  & 0.084 & 14.1 & 1.1\\
  118 & 56.1186 & 32.1230 & 2MASS J03442847+3207224 & 76 & 66 & 7.01 & 0.07 & 7 & 0.17 & 14.2 & 1.3\\
  125 & 56.1602 & 32.1266 & 2MASS J03443845+3207356 & 104 & 36 & 5.21 & 0.05 & 5.1 & 0.27 & 14.2 & 1.4\\
  126 & 55.5462 & 32.3783 & 2MASS J03421109+3222418 &  &  & 11.0 & 0.1 &  & 0.05 & 14.2 & 1.1\\
  139 & 56.4865 & 32.1620 & 2MASS J03455676+3209428 &  &  & 1.26 & 0.01 &  & 0.031 & 14.4 & 0.7\\
  145 & 55.8836 & 32.1049 & 2MASS J03433205+3206172 &  & 94 & 5.49 & 0.05 &  & 0.18 & 14.5 & 1.2\\
  147 & 56.1672 & 32.1928 & 2MASS J03444011+3211341 &  & 59 & 2.14 & 0.02 &  & 0.10 & 14.6 & 1.4\\
  151 & 56.0007 & 32.3708 & 2MASS J03440015+3222144 &  &  & 12.15 & 0.12 &  & 0.028 & 14.5 & 0.9\\
  154 & 56.2572 & 32.2411 & 2MASS J03450174+3214276 & 134 & 39 & 16.4 & 0.2 & 16.4 & 0.10 & 14.6 & 1.5\\
  157 & 56.1066 & 32.2084 & 2MASS J03442557+3212299 & 68 & 64 & 8.37 & 0.08 & 8.4 & 0.47 & 14.8 & 1.4\\
  158 & 56.4391 & 32.1445 & 2MASS J03454539+3208401 &  &  & 11.91 & 0.12 &  & 0.052 & 14.7 & 0.8\\
  162 & 55.7459 & 32.5015 & 2MASS J03425901+3230053 &  &  & 21.0 & 0.2 &  & 0.054 & 14.6 & 1.4\\
  163 & 56.1126 & 32.0789 & 2MASS J03442702+3204436 & 71 & 69 & 8.93 & 0.09 & 9.1 & 0.20 & 14.7 & 1.2\\
  168 & 55.6818 & 31.9876 & 2MASS J03424360+3159150 & 10 &  & 2.18 & 0.02 & 2.2 & 0.094 & 14.8 & 1.4\\
  174 & 56.1775 & 32.1055 & 2MASS J03444261+3206194 & 121 & 146 & 11.5 & 0.1 & 12 & 0.16 & 14.8 & 1.2\\
  176 & 56.1558 & 32.1033 & 2MASS J03443740+3206118 & 97 & 82 & 6.28 & 0.06 & 6.2 & 0.095 & 15.0 & 1.4\\
  184 & 56.3166 & 32.5146 & 2MASS J03451598+3230519 & 135 &  & 4.04 & 0.04 & 4.1 & 0.12 & 14.9 & 1.4\\
  187 & 55.9618 & 31.9059 & 2MASS J03435084+3154210 &  &  & 4.04 & 0.04 &  & 0.12 & 14.9 & 1.3\\
  190 & 56.4079 & 32.1403 & 2MASS J03453789+3208249 &  &  & 0.28 & 0.01 &  & 0.012 & 14.9 & 1.0\\
  195 & 55.6255 & 31.7028 & 2MASS J03423010+3142104 &  &  & 29.9 & 0.3 &  & 0.088 & 15.0 & 1.2\\
  213 & 56.4741 & 32.4451 & 2MASS J03455377+3226418 &  &  & 0.434 & 0.002 &  & 0.075 & 15.0 & 0.9\\
  217 & 55.6693 & 32.2859 & 2MASS J03424056+3217101 &  &  & 3.34 & 0.03 &  & 0.085 & 15.3 & 1.2\\
  223 & 55.9531 & 32.1260 & 2MASS J03434875+3207332 & 25 & 111 & 39.1 & 0.4 & 20.3 & 0.07 & 15.3 & 1.3\\
  226 & 56.0091 & 32.3278 & 2MASS J03440216+3219399 & 35 & 118 & 6.11 & 0.06 & 6.2 & 0.35 & 15.2 & 1.5\\
  245 & 56.3181 & 32.1056 & 2MASS J03451634+3206199 &  & 1933 & 28.5 & 0.3 &  & 1.9 & 15.8 & 1.5\\
  246 & 56.3474 & 32.4104 & 2MASS J03452338+3224369 &  &  & 8.69 & 0.09 &  & 0.097 & 15.4 & 1.4\\
  251 & 56.0746 & 32.2057 & 2MASS J03441791+3212203 & 46 & 93 & 4.39 & 0.04 & 4.5 & 0.16 & 15.3 & 1.4\\
  252 & 55.9989 & 32.2342 & 2MASS J03435970+3214028 & 34 & 87 & 14.04 & 0.1 & 14 & 0.16 & 15.3 & 1.6\\
  260 & 55.9643 & 32.5303 & 2MASS J03435141+3231486 & 28 &  & 3.75 & 0.04 & 3.8 & 0.13 & 15.4 & 1.6\\
  262 & 56.0476 & 32.3279 & 2MASS J03441143+3219401 & 41 & 137 & 0.766 & 0.008 & 0.8 & 0.092 & 15.6 & 1.4\\
  268 & 56.0469 & 32.1034 & 2MASS J03441125+3206121 & 40 & 105 & 10.5 & 0.1 & 9.1 & 0.22 & 15.5 & 1.5\\
  277 & 55.9843 & 32.5050 & 2MASS J03435622+3230178 & 30 &  & 8.23 & 0.08 & 8.1 & 0.10 & 15.5 & 1.6\\
  279 & 55.6486 & 32.2889 & 2MASS J03423566+3217198 & 7 &  & 1.15 & 0.02 & 1.2 & 0.035 & 15.5 & 0.9\\
  280 & 56.1172 & 32.2667 & 2MASS J03442812+3216002 & 75 & 85 & 2.67 & 0.03 & 2.7 & 0.22 & 15.9 & 1.2\\
  282 & 56.1162 & 32.1255 & 2MASS J03442787+3207316 & 74 & 86 & 6.55 & 0.07 & 5.4 & 0.084 & 15.6 & 1.5\\
  292 & 55.7311 & 31.9535 & 2MASS J03425546+3157123 & 11 &  & 0.6508 & 0.001 & 1.9 & 0.12 & 15.5 & 1.2\\
\hline
	\end{tabular}
\end{table*}
\begin{table*}
	\contcaption{}
\begin{tabular}{rrrlrrrrrrrr}
	\hline
	ID & RA & DEC & Identifier & CB & LRL & $P$ & $\Delta P$ & $P_\mathrm{CB}$ & Amplitude & $R$ & $(V-R)$\\
	& [\degr] & [\degr] & & & & [d] & [d] & [d] & [mag] & [mag] & [mag] \\
	\hline
  296 & 56.1229 & 32.4403 & 2MASS J03442949+3226250 &  &  & 9.5 & 0.1 &  & 0.044 & 15.7 & 1.3\\
  297 & 56.1605 & 32.1336 & 2MASS J03443854+3208006 & 105 & 58 & 7.69 & 0.08 & 7.3 & 0.20 & 15.7 & 1.7\\
  304 & 56.0986 & 32.1130 & 2MASS J03442366+3206465 & 64 & 92 & 9.7 & 0.1 & 10 & 0.28 & 15.6 & 1.5\\
  305 & 56.1605 & 32.2167 & 2MASS J03443838+3212597 & 103 & 144 & 13.5 & 0.1 & 13.5 & 0.27 & 15.8 & 1.5\\
  317 & 55.9842 & 32.1434 & 2MASS J03435619+3208362 &  & 142 & 3.91 & 0.04 &  & 0.066 & 15.9 & 1.4\\
  318 & 55.9495 & 32.2992 & 2MASS J03434788+3217567 & 24 & 126 & 9.68 & 0.1 & 9.8 & 0.16 & 15.7 & 1.5\\
  327 & 56.3845 & 32.0542 & 2MASS J03453230+3203150 & 140 & 10289 & 0.705 & 0.001 & 0.7 & 0.17 & 15.8 & 1.2\\
  341 & 56.0901 & 32.1772 & 2MASS J03442161+3210376 & 58 & 41 & 7.56 & 0.08 & 2.8 & 1.3 & 16.6 & 2.1\\
  346 & 56.0957 & 32.1993 & 2MASS J03442297+3211572 &  & 120 & 5.06 & 0.05 &  & 0.16 & 15.9 & 1.5\\
  347 & 56.1188 & 31.9984 & 2MASS J03442851+3159539 &  & 68 & 26.9 & 0.3 &  & 0.22 & 15.8 & 1.5\\
  348 & 56.1384 & 32.2581 & 2MASS J03443321+3215290 &  & 122 & 6.96 & 0.07 &  & 0.098 & 15.7 & 1.7\\
  349 & 56.0930 & 32.2002 & 2MASS J03442232+3212007 & 61 & 100 & 0.83 & 0.01 & 8.4 & 0.099 & 16.1 & 1.6\\
  353 & 55.6923 & 32.4549 & 2MASS J03424614+3227172 &  &  & 12.1 & 0.1 &  & 0.075 & 16.0 & 1.0\\
  355 & 56.1111 & 32.0662 & 2MASS J03442663+3203583 & 69 & 62 & 3.09 & 0.03 & 3.1 & 0.16 & 16.0 & 1.8\\
  357 & 56.2907 & 32.2268 & 2MASS J03450986+3213350 &  &  & 2.47 & 0.02 &  & 0.03 & 16.0 & 1.1\\
  364 & 56.0170 & 32.1214 & 2MASS J03440410+3207170 &  & 174 & 10.0 & 0.1 &  & 0.20 & 16.3 & 1.6\\
  366 & 56.5065 & 32.1606 & 2MASS J03460161+3209375 &  &  & 1.67 & 0.02 &  & 0.087 & 16.1 & 0.9\\
  384 & 56.1064 & 32.1920 & 2MASS J03442555+3211307 & 67 & 60 B & 4.86 & 0.05 & 5.4 & 0.36 & 16.3 & 1.6\\
  386 & 55.7794 & 32.1718 & 2MASS J03430704+3210182 & 14 &  & 6.14 & 0.06 & 6.2 & 0.19 & 16.1 & 1.7\\
  387 & 56.1859 & 32.1369 & 2MASS J03444458+3208125 &  & 103 & 25.7 & 0.3 &  & 0.89 & 17.0 & 2.0\\
  391 & 56.0898 & 32.1716 & 2MASS J03442155+3210174 & 56 & 116 & 7.09 & 0.07 & 7 & 0.36 & 16.1 & 1.8\\
  394 & 55.9558 & 32.1778 & 2MASS J03434939+3210398 & 27 & 147 & 6.4 & 0.06 & 13.1 & 0.19 & 16.2 & 1.4\\
  407 & 56.2123 & 32.2693 & 2MASS J03445096+3216093 & 128 & 101 & 12.1 & 0.1 & 13.1 & 0.27 & 16.0 & 1.5\\
  421 & 56.0929 & 32.0953 & 2MASS J03442228+3205427 & 60 & 61 & 11.9 & 0.1 & 30 & 0.50 & 16.5 & 1.3\\
  441 & 55.7352 & 31.7106 & 2MASS J03425646+3142382 &  &  & 8.62 & 0.09 &  & 0.23 & 16.2 & 2.1\\
  444 & 55.5361 & 32.4745 & 2MASS J03420867+3228276 &  &  & 2.84 & 0.03 &  & 0.14 & 16.3 & 1.7\\
  465 & 56.0897 & 32.2528 & 2MASS J03442156+3215098 &  & 185 & 8.17 & 0.08 &  & 0.31 & 16.9 & 0.4\\
  470 & 56.2338 & 32.0991 & 2MASS J03445611+3205564 & 131 & 188 & 3.31 & 0.03 & 3.3 & 0.087 & 16.3 & 1.3\\
  476 & 56.3981 & 31.9406 & 2MASS J03453551+3156257 & 141 &  & 9.28 & 0.09 & 9.6 & 0.13 & 16.4 & 1.5\\
  477 & 56.1691 & 32.3865 & 2MASS J03444061+3223110 & 114 &  & 4.03 & 0.04 & 4.1 & 0.14 & 16.3 & 1.5\\
  502 & 56.1023 & 32.0659 & 2MASS J03442457+3203571 & 65 & 123 & 4.94 & 0.05 & 4.9 & 0.33 & 17.1 & 1.9\\
  505 & 56.2034 & 32.2228 & 2MASS J03444881+3213218 & 127 & 178 & 6.9 & 0.07 & 6.9 & 0.25 & 16.5 & 1.7\\
  509 & 56.1812 & 32.1286 & 2MASS J03444351+3207427 &  & 52 & 8.14 & 0.08 &  & 0.13 & 16.6 & 1.9\\
  514 & 56.2198 & 32.0158 & 2MASS J03445274+3200565 &  & 1939 & 10.9 & 0.1 &  & 0.15 & 16.7 & 1.6\\
  523 & 55.9983 & 32.2654 & 2MASS J03435953+3215551 & 33 & 155 & 7.59 & 0.08 & 7.6 & 0.33 & 16.7 & 1.4\\
  525 & 56.3444 & 31.7239 & 2MASS J03452267+3143259 &  &  & 17.1 & 0.2 &  & 0.15 & 16.4 & 1.4\\
  526 & 55.6077 & 32.3507 & 2MASS J03422585+3221022 & 2 &  & 17.3 & 0.2 & 17.7 & 0.22 & 16.5 & 1.6\\
  543 & 55.6177 & 32.5133 & 2MASS J03422824+3230479 & 3 &  & 4.34 & 0.04 & 4.3 & 0.45 & 16.3 & 1.5\\
  544 & 55.9534 & 32.2644 & 2MASS J03434881+3215515 & 26 & 162 & 2.88 & 0.03 & 2.9 & 0.18 & 16.9 & 1.7\\
  565 & 55.5846 & 32.0920 & 2MASS J03422033+3205310 & 1 &  & 7.25 & 0.07 & 7.7 & 0.57 & 16.6 & 1.8\\
  570 & 56.0802 & 32.1263 & PSZ2003 J034419.2+320734 & 49 & 99 A & 7.55 & 0.08 & 7.6 & 1.3 & 16.0 & 1.5\\
  596 & 56.3977 & 32.0572 & 2MASS J03453545+3203259 &  & 10284 & 1.57 & 0.02 &  & 0.26 & 16.6 & 1.6\\
  607 & 56.2081 & 32.0628 & 2MASS J03444998+3203455 &  &  & 220 & 2 &  & 0.085 & 16.7 & 1.7\\
  608 & 56.0906 & 32.2087 & 2MASS J03442176+3212312 &  & 180 & 3.26 & 0.03 &  & 0.08 & 16.7 & 1.7\\
  611 & 56.1249 & 32.3230 & 2MASS J03442997+3219227 &  & 104 & 8.67 & 0.09 &  & 0.25 & 16.8 & 2.3\\
  616 & 55.6342 & 31.7272 & 2MASS J03423219+3143382 &  &  & 12.2 & 0.1 &  & 0.21 & 16.7 & 1.9\\
  617 & 55.6703 & 32.2264 & 2MASS J03424086+3213347 & 9 &  & 0.47375 & 0.00025 & 0.5 & 0.085 & 16.6 & 1.8\\
  640 & 56.1495 & 32.2645 & 2MASS J03443588+3215533 &  & 181 & 7.7 & 0.08 &  & 0.17 & 16.6 & 1.7\\
  670 & 56.3423 & 32.0345 & 2MASS J03452214+3202040 &  & 10373 & 0.78 & 0.01 &  & 0.10 & 16.9 & 1.3\\
  683 & 56.0756 & 32.0825 & 2MASS J03441816+3204570 &  & 31 & 3.34 & 0.03 &  & 0.14 & 17.4 & 2.1\\
  696 & 55.6000 & 31.9005 & 2MASS J03422398+3154016 &  &  & 0.579 & 0.001 &  & 0.07 & 17.0 & 1.3\\
  699 & 56.2718 & 32.1653 & 2MASS J03450521+3209544 &  & 177 & 4.95 & 0.05 &  & 0.16 & 17.1 & 1.6\\
  701 & 56.3406 & 32.5352 & 2MASS J03452174+3232065 &  &  & 5.07 & 0.05 &  & 0.2 & 16.9 & 1.9\\
  723 & 56.0283 & 32.1318 & 2MASS J03440678+3207540 & 37 & 156 & 2.61 & 0.03 & 1.3 & 0.13 & 17.1 & 2.0\\
  735 & 55.8874 & 32.4675 & 2MASS J03433299+3228027 & 22 &  & 11.7 & 0.1 & 12.2 & 0.23 & 17.3 & 2.0\\
  758 & 56.2620 & 32.1169 & 2MASS J03450285+3207006 &  & 150 & 2.39 & 0.02 &  & 0.14 & 17.3 & 1.7\\
  774 & 56.3116 & 32.4332 & 2MASS J03451480+3225594 &  &  & 5.58 & 0.06 &  & 0.2 & 17.0 & 1.9\\
  775 & 56.1584 & 32.1937 & 2MASS J03443800+3211370 &  & 193 & 2.39 & 0.02 &  & 0.25 & 17.4 & 1.4\\
  777 & 56.4146 & 31.7148 & 2MASS J03453948+3142528 &  &  & 0.4882 & 0.0015 &  & 0.27 & 17.3 & 0.9\\
  796 & 56.1134 & 32.2393 & 2MASS J03442724+3214209 &  & 132 & 4.4 & 0.04 &  & 1.2 & 17.2 & 1.7\\
  798 & 56.4449 & 32.4804 & 2MASS J03454675+3228487 & 143 &  & 10.33 & 0.1 & 10.5 & 0.20 & 17 & 1.7\\
  800 & 56.1874 & 32.2268 & 2MASS J03444495+3213364 &  & 112 & 20.0 & 0.2 &  & 0.20 & 17.1 & 1.8\\
  \hline

	\end{tabular}
\end{table*}
\begin{table*}
	\contcaption{}
\begin{tabular}{rrrlrrrrrrrr}
	\hline
	ID & RA & DEC & Identifier & CB & LRL & $P$ & $\Delta P$ & $P_\mathrm{CB}$ & Amplitude & $R$ & $(V-R)$\\
	& [\degr] & [\degr] & & & & [d] & [d] & [d] & [mag] & [mag] & [mag] \\
	\hline
  807 & 55.6372 & 31.7057 & 2MASS J03423291+3142205 &  &  & 10.69 & 0.11 &  & 1.0 & 16.9 & 2.1\\
  811 & 56.1616 & 32.3183 & 2MASS J03443878+3219056 & 108 & 208 & 4.85 & 0.05 & 4.8 & 0.20 & 16.9 & 2.4\\
  829 & 56.0312 & 32.0691 & 2MASS J03440750+3204088 &  & 214 & 2.34 & 0.02 &  & 0.24 & 17.5 & 1.6\\
  848 & 55.7330 & 31.7464 & 2MASS J03425593+3144463 &  &  & 2.46 & 0.02 &  & 0.12 & 17.1 & 1.6\\
  849 & 56.1574 & 32.2050 & 2MASS J03443777+3212181 & 100 & 154 & 3.25 & 0.03 & 3.3 & 0.25 & 17.4 & 1.7\\
  898 & 56.4328 & 32.4098 & 2MASS J03454385+3224350 & 142 &  & 14.4 & 0.1 & 14.3 & 0.30 & 17.5 & 1.8\\
  927 & 56.4929 & 32.4466 & 2MASS J03455824+3226475 &  &  & 7.72 & 0.07 &  & 0.80 & 18.2 & 1.1\\
  935 & 55.9526 & 32.2307 & 2MASS J03434862+3213507 &  & 261 & 0.638 & 0.006 &  & 0.30 & 18.0 & 1.3\\
  941 & 56.0191 & 32.4684 & 2MASS J03440459+3228062 &  &  & 6.96 & 0.07 &  & 0.40 & 17.5 & 1.7\\
  952 & 56.0834 & 32.1127 & 2MASS J03442001+3206455 & 50 & 210 & 8.62 & 0.09 & 8.6 & 0.45 & 17.6 & 1.4\\
  955 & 56.1240 & 32.1777 & 2MASS J03442972+3210398 &  & 40 & 4.6 & 0.05 &  & 0.25 & 15.4 & 1.4\\
  956 & 56.1363 & 32.1545 & 2MASS J03443276+3209157 & 82 & 88 & 5.28 & 0.05 & 5.5 & 0.33 & 16.2 & 0.7\\
  957 & 56.1355 & 32.1490 & 2MASS J03443257+3208558 & 80 & 71 & 1.82 & 0.02 & 6.7 & 0.14 & 16.1 & 2.1\\
  958 & 56.1311 & 32.1457 & 2MASS J03443153+3208449 & 78 & 29 & 2.22 & 0.02 & 10.8 & 0.069 & 12.8 & 1.1\\
  959 & 56.1363 & 32.1438 & 2MASS J03443274+3208374 & 81 & 16 & 2.62 & 0.03 & 2.6 & 0.088 & 12.5 & 1.0\\
  960 & 56.1611 & 32.1450 & 2MASS J03443871+3208420 & 107 & 23 & 2.41 & 0.02 & 2.4 & 0.12 & 15.4 & 1.6\\
  961 & 56.1614 & 32.1489 & 2MASS J03443869+3208567 & 106 & 108 & 2.79 & 0.03 & 2.1 & 0.13 & 16.0 & 1.6\\
  962 & 56.1558 & 32.1504 & 2MASS J03443741+3209009 & 99 & 83 & 8.34 & 0.08 & 8.4 & 0.43 & 16.5 & 0.9\\
  964 & 56.1633 & 32.1623 & 2MASS J03443919+3209448 & 110 & 91 & 3.98 & 0.04 & 3.9 & 0.54 & 16.3 & 1.4\\
  965 & 56.1416 & 32.1482 & 2MASS J03443398+3208541 & 86 & 65 & 15.8 & 0.2 & 16.4 & 0.21 & 14.7 & 1.5\\
  967 & 56.0903 & 32.1070 & 2MASS J03442166+3206248 & 57 & 125 & 8.36 & 0.08 & 8.4 & 0.18 &  & \\
  968 & 56.0939 & 32.0314 & 2MASS J03442257+3201536 & 62 & 72 & 44.7 & 0.4 & 1 & 0.21 & 15.9 & 1.6\\
  972 & 56.1083 & 32.2750 & 2MASS J03442595+3216306 &  & (972) & 7.56 & 0.08 &  & 0.032 & 14.2 & 0.8\\
  973 & 56.0406 & 32.2868 & 2MASS J03440973+3217130 &  &  & 9.58 & 0.1 &  & 0.2 & 15.5 & 0.5\\
  974 & 55.6466 & 32.5650 & 2MASS J03423520+3233544 & 6 &  & 0.422 & 0.002 & 0.2 & 0.25 &  & \\
  975 & 55.8014 & 32.5698 & 2MASS J03431233+3234114 & 17 &  & 0.446 & 0.002 & 0.20 & 0.35 &  & \\
  996 & 56.1153 & 32.5637 & 2MASS J03442766+3233495 & 73 &  & 1.59 & 0.02 & 2.6 & 0.094 & 15.8 & 1.5\\
  998 & 56.0927 & 32.5625 & 2MASS J03442225+3233449 &  &  & 12.13 & 0.12 &  & 0.043 & 14.5 & 1.0\\
  1001 & 56.4838 & 32.572 & 2MASS J03455608+3234190 &  &  & 11.88 & 0.12 &  & 0.13 & 15.8 & 1.2\\
  \hline
\end{tabular}
\end{table*}

\begin{figure*}
	\includegraphics[width=\textwidth]{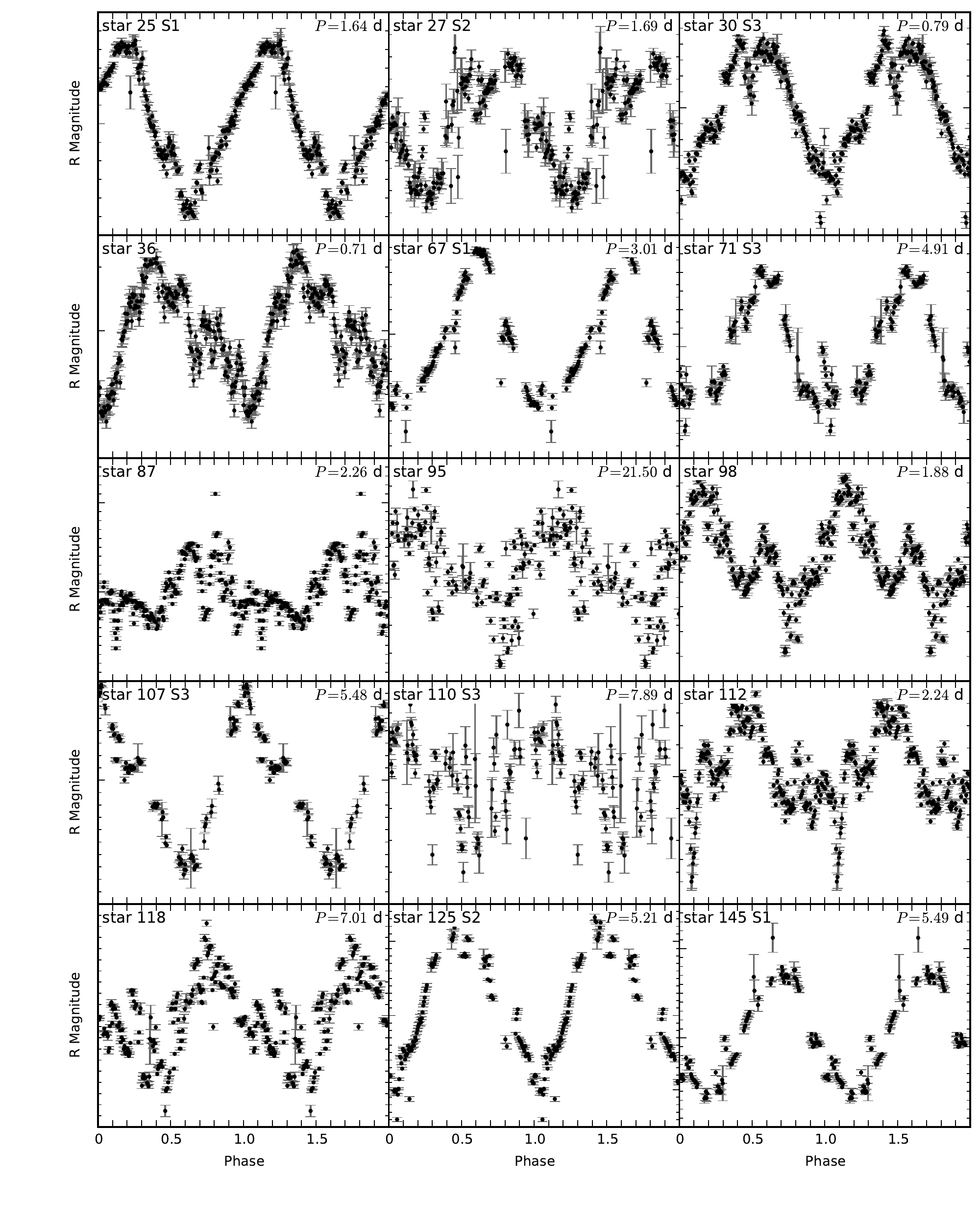}
	\caption{Phase-folded and binned (200 bins) light curves of all periodically variable IC~348 members. We present the double phase for easier visualization (data points in the phase between 1 and 2 are repeated from phase 0 to 1). For stars without a coherent phase over the full time span of the observations only one season is shown. This is indicated by S$n$ added to the star number, where $n$ is an integer from 1 to 3 (compare \autoref{tab:observations}). The spacing is 0.01\,mag for minor ticks and 0.1\,mag for major ticks on the magnitude axis.}
	\label{fig:lcs}
\end{figure*}

\begin{figure*}
	\includegraphics[width=\textwidth]{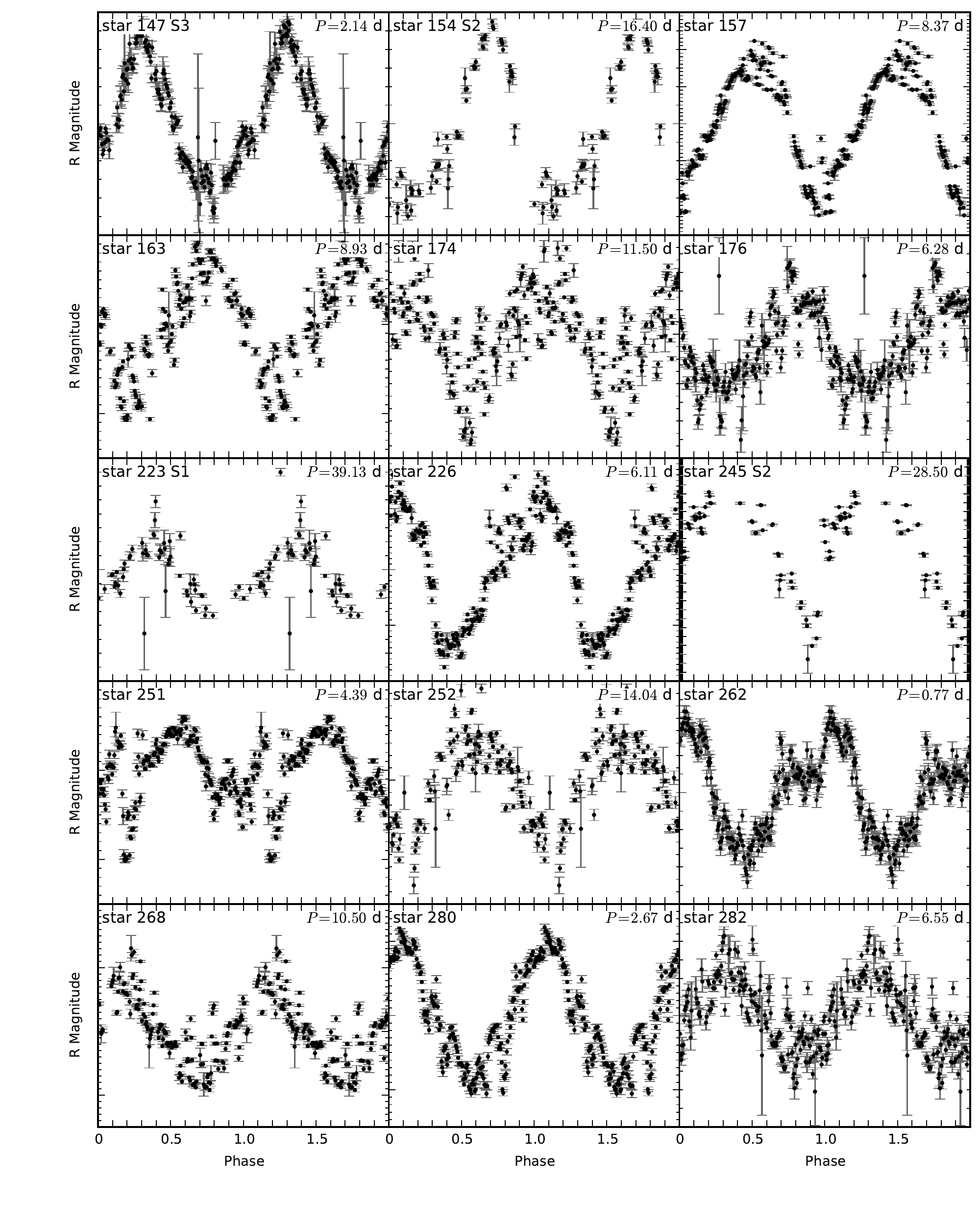}
	\contcaption{}
\end{figure*}

\begin{figure*}
	\includegraphics[width=\textwidth]{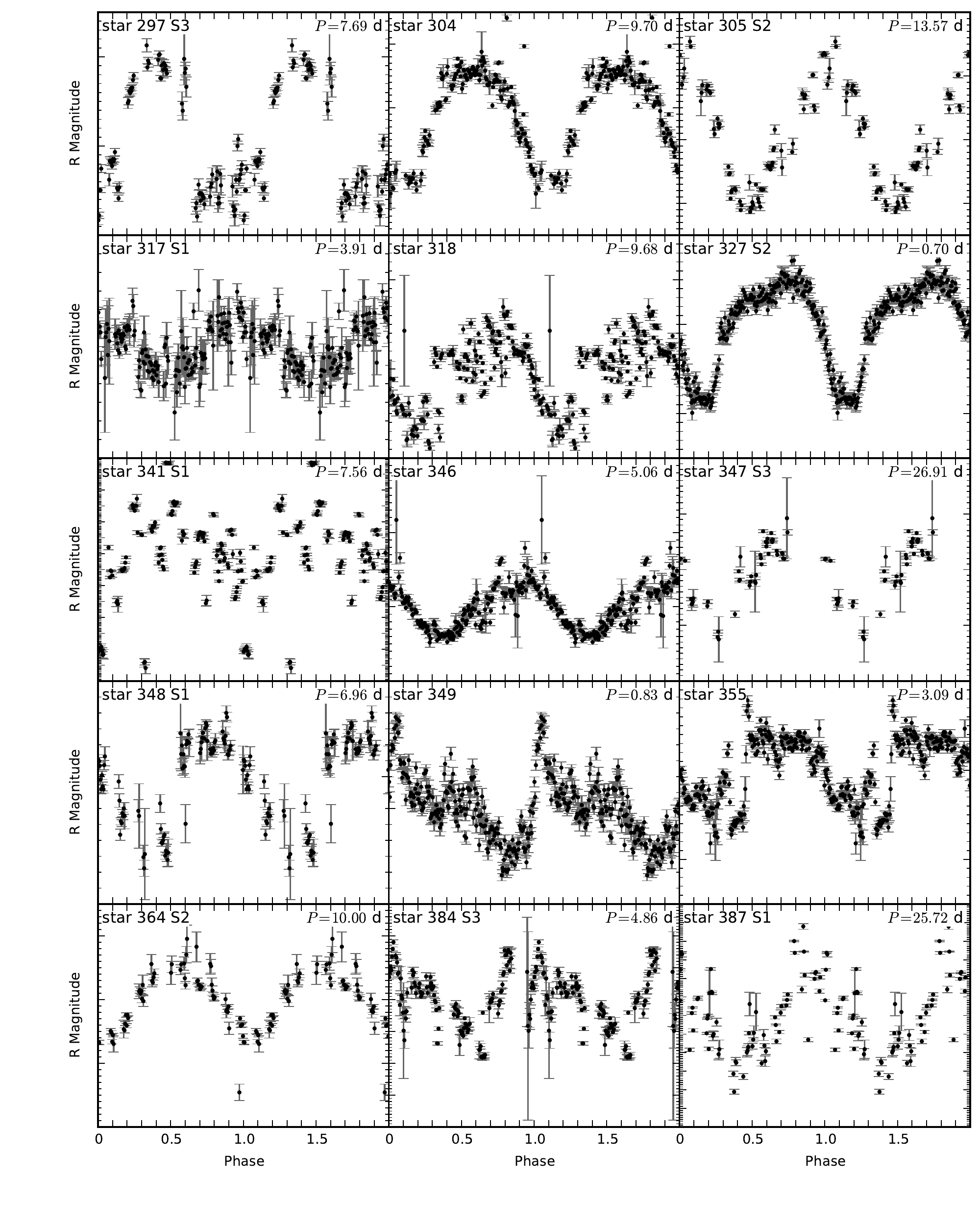}
	\contcaption{}
\end{figure*}

\begin{figure*}
	\includegraphics[width=\textwidth]{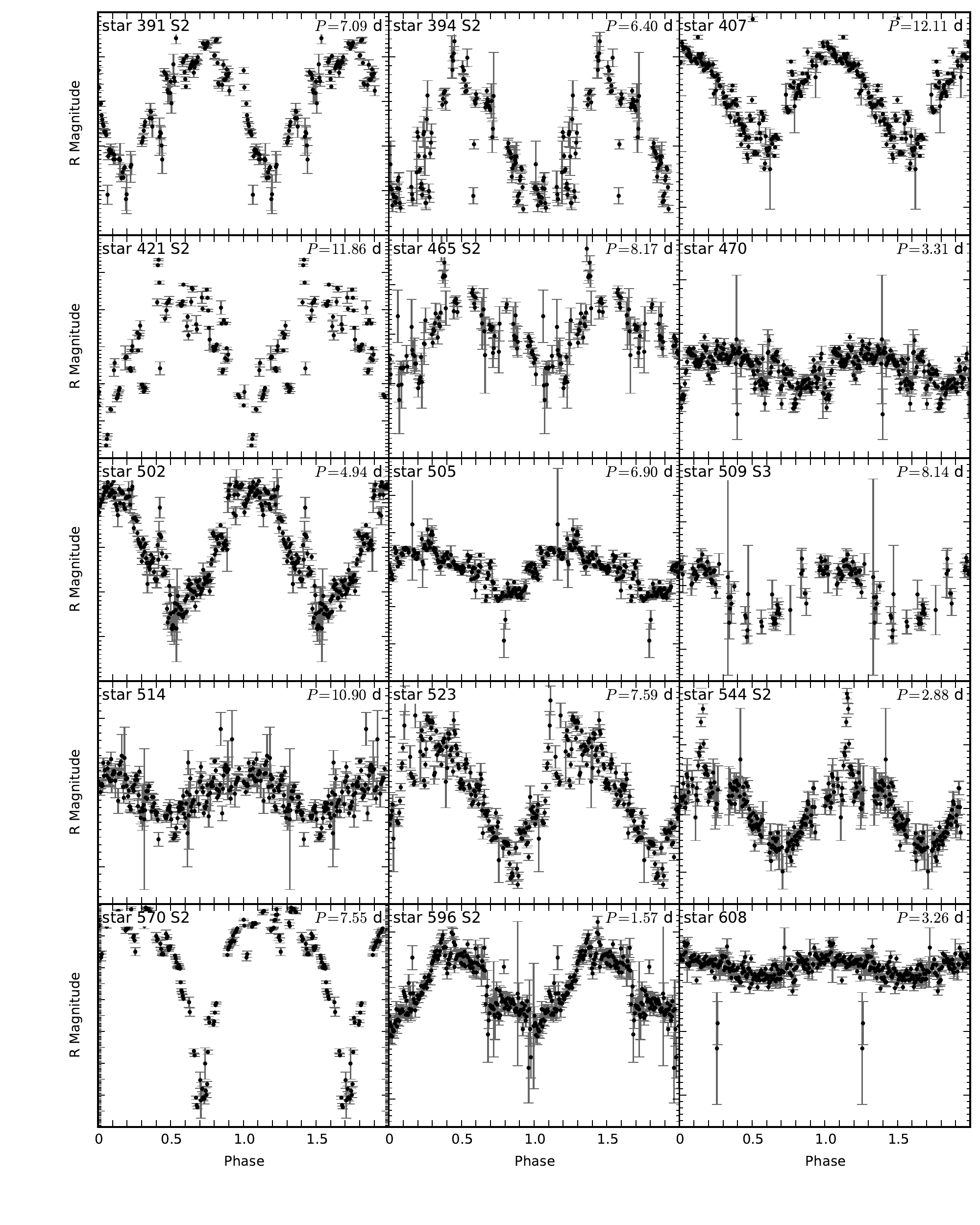}
	\contcaption{}
\end{figure*}

\begin{figure*}
	\includegraphics[width=\textwidth]{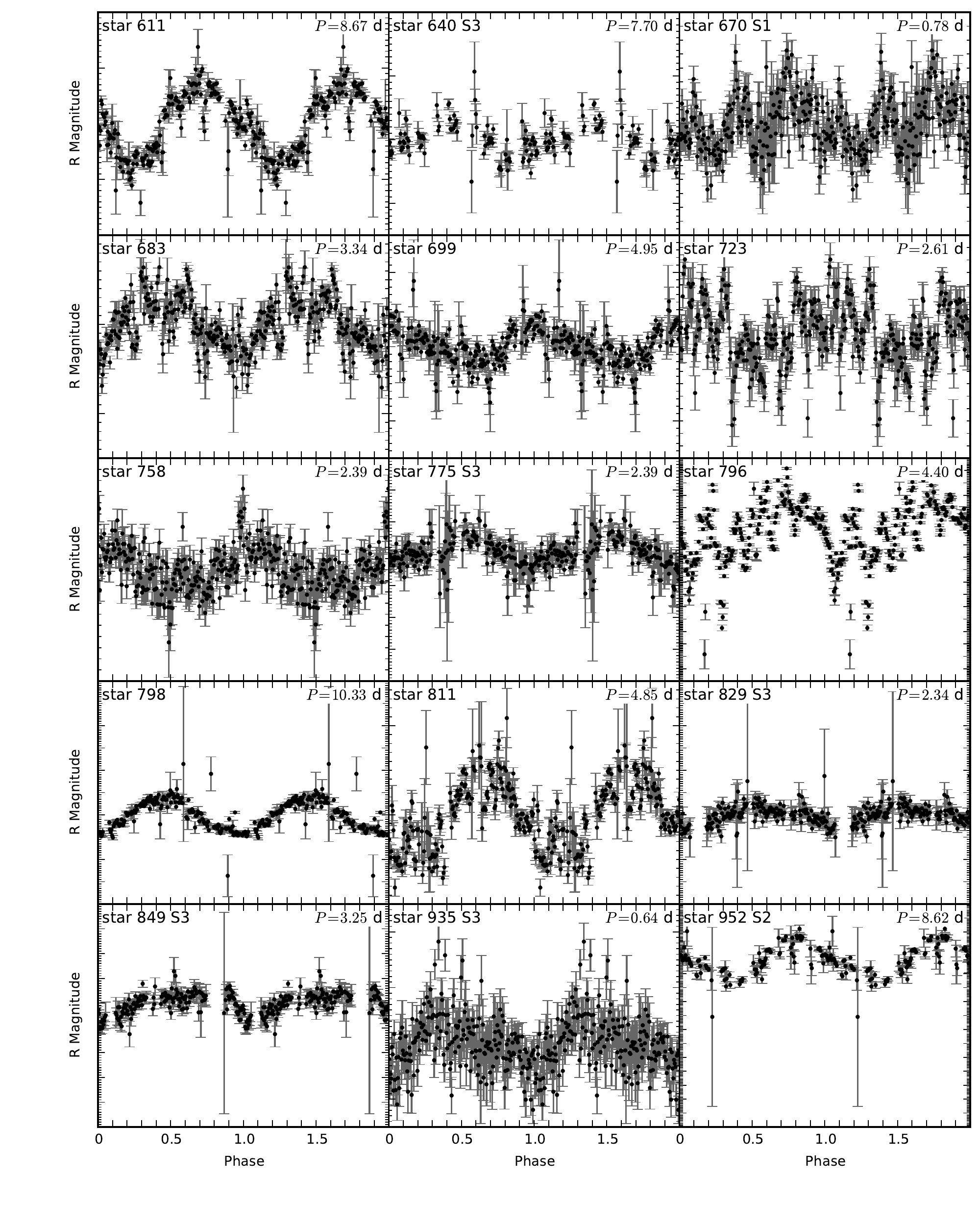}
	\contcaption{}
\end{figure*}

\begin{figure*}
	\includegraphics[width=\textwidth]{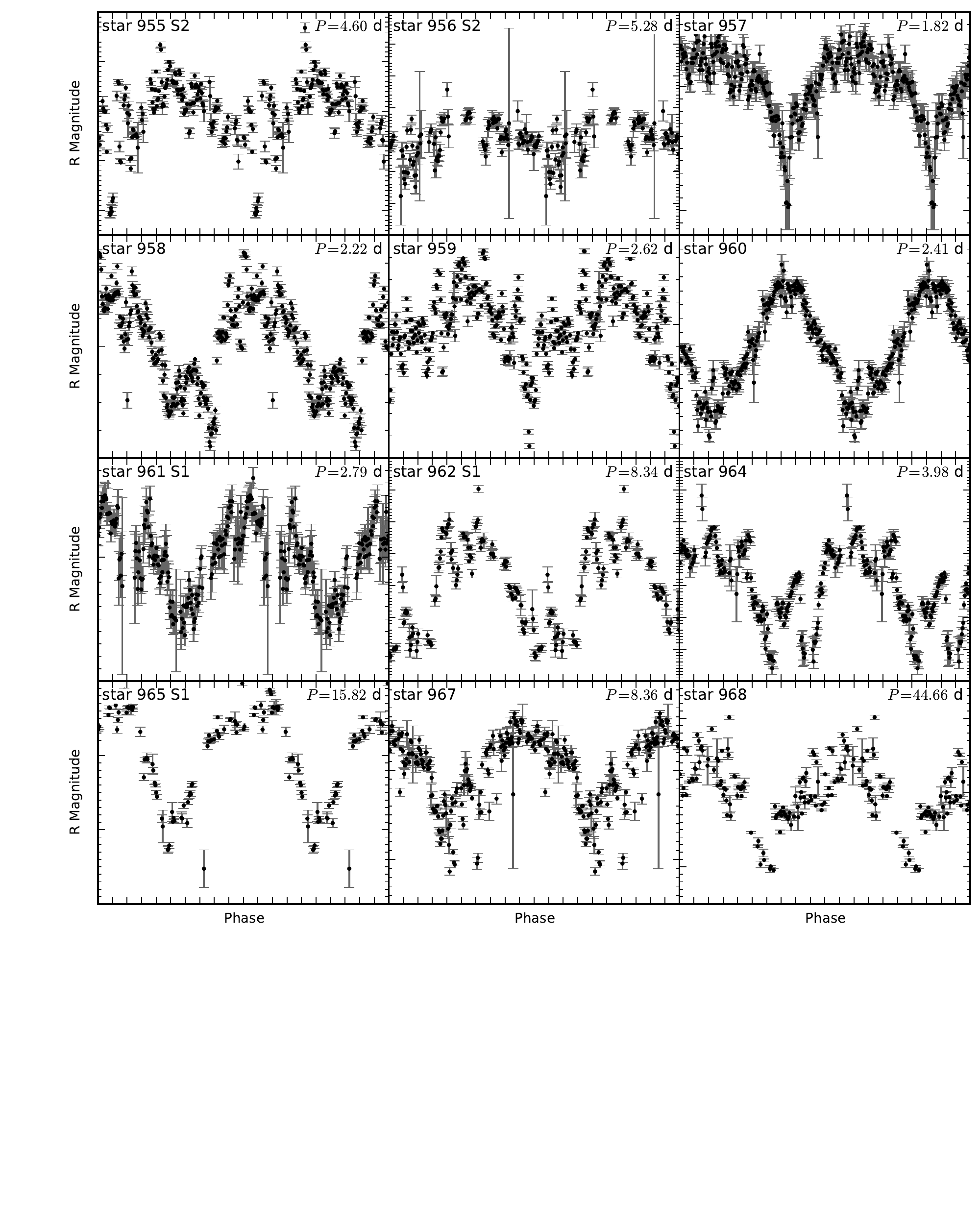}
	\contcaption{}
\end{figure*}

%%%%%%%%%%%%%%%%%%%%%%%%%%%%%%%%%%%%%%%%%%%%%%%%%%

% Don't change these lines
\bsp	% typesetting comment
\label{lastpage}
\end{document}